\documentclass[aps,groupedaddress,showkeys,nofootinbib]{revtex4-1}
\usepackage[T1]{fontenc}
\usepackage{ae,aecompl}
\usepackage{graphicx}	
\usepackage{amsmath}	
\usepackage{amssymb}	
\usepackage{multirow}
\usepackage{subfigure}
\usepackage{stackrel}
\usepackage{float}
\usepackage{natbib}
\usepackage[unicode=true]{hyperref}

\begin{document}
\title{Addressing $\gamma$-ray 
emissions from dark matter annihilations in 45 milky way satellite galaxies 
and in extragalactic sources with particle dark matter models}

\author{Ashadul Halder}
\email{ashadul.halder@gmail.com}
\affiliation{Department of Physics, St. Xavier's College, 30, Mother Teresa Sarani, Kolkata-700016, India.}

\author{Shibaji Banerjee}
\email{shiva@sxccal.edu}
\affiliation{Department of Physics, St. Xavier's College, 30, Mother Teresa Sarani, Kolkata-700016, India.}

\author{Madhurima Pandey}
\email{madhurima.pandey@saha.ac.in}
\affiliation{Astroparticle Physics and Cosmology Division, Saha Institute of Nuclear Physics, HBNI, 1/AF Bidhannagar, Kolkata 700064, India}

\author{Debasish Majumdar}
\email{debasish.majumdar@saha.ac.in}
\affiliation{Astroparticle Physics and Cosmology Division, Saha Institute of Nuclear Physics, HBNI, 1/AF Bidhannagar, Kolkata 700064, India}

\begin{abstract}
The mass to luminosity ratio of the dwarf satellite galaxies in the Milky Way suggests that these dwarf galaxies may contain substantial 
dark matter. The dark matter at the dense region such as within or at the 
vicinity of the centres of these dwarf galaxies may undergo the process of 
self annihilation and produce $\gamma$-rays as the end product. The satellite borne $\gamma$-ray telescope such as {\it Fermi}-LAT reported the detection of $\gamma$-rays from around 45 Dwarf Spheroidals (dSphs) of Milky Way. In this work, we consider particle dark matter models described in the literature and 
after studying their phenomenologies, we calculate the $\gamma$-ray 
fluxes from the self annihilation of the dark matter within the framework of these models in case of each of these 45 dSphs. we then compare the 
computed results with the observational upper bounds for $\gamma$-ray flux 
reported by {\it Fermi}-LAT and Dark Energy Survey (DES) for each of the 45 dSphs. 
The fluxes are calculated by adopting different dark matter density profiles. 
We then extend similar analysis for 
the observational upper bounds given by {\it Fermi}-LAT for the continuum $\gamma$-ray fluxes originating from extragalactic sources.
\end{abstract}

\maketitle


\section{Introduction} \label{sec:intro}

Although the existence of dark matter (DM) in the Universe is now well 
established, any direct signature of the dark matter is still eluding the worldwide
endeavours at different direct dark matter search experiments. The indirect 
search for dark matter involves detection of the known Standard Model (SM)
particles that can be produced by possible dark matter annihilation 
(or decay) in cosmos.
Although the cosmic relics, the dark matter can undergo self annihilation 
if it is accumulated in considerable magnitude by being captured, under the 
influence of gravity, inside massive astrophysical bodies. 
In literature there are indications that the emissions of excess 
$\gamma$-rays from the Galactic Centre (GC) region (detected by {\it Fermi}-LAT 
satellite borne experiment) could have been originated from the 
annihilation of dark matter at GC region. 
The dwarf spheroidals are the satellite galaxies to the Milky 
Way and they fail to grow as matured galaxies. These dwarf spheroidal
galaxies (dSphs) are generally of low luminosities and contain population of 
older stars with little dust. The dwarf spheroidals could be very rich 
in dark matter. These galaxies would have been tidally disrupted
but the presence of dark matter provides the necessary gravitational
pull. The existence of dark matter in dwarf spheroidals can also be realised
by studying their mass to luminosity ratios. From several observations 
the estimated mass to luminosity ratios $(M/L)$ are found to be much more 
than the same for the sun $\left(\left | \frac {M} {L} \right |\right)$. 
The dark matter at dSphs can undergo self annihilation and 
produce $\gamma$-rays. 

The {\it Fermi}-LAT satellite borne observations and Dark Energy Survey (DES) 
have reported the 
upper bounds of the $\gamma$-ray spectra for several dwarf 
galaxies \cite{fermilat,Fermi-LAT:2016uux}. Here 
in this work, we consider two particle dark matter models 
(one is the simple extension of Standard Model and the other is inspired by a
Beyond Standard Model (BSM) theory of particle physics) 
and for the 
dark matter candidates in each of these two models, we compute 
the expected $\gamma$-ray flux from all the 45 dSphs mentioned 
above by considering the dark matter annihilations at those dSphs.
These computed results are then compared with the
observational upper bounds for $\gamma$-ray flux for each of the 45 dSphs. 
The {\it Fermi}-LAT experiment also provides \cite{2010PhRvL.104j1101A,Ackermann:2014usa} 
the observational results for extragalactic $\gamma$-ray flux. 
If the $\gamma$-rays (or a component of that) also originate from dark matter 
annihilation, then this too could be indirect signal of dark matter. 
The dark matter annihilation to $\gamma$-rays in extragalactic source 
is considered to add to the extragalactic $\gamma$-ray flux. To this end, 
we compute the extragalactic $\gamma$-ray flux from other possible sources such as GRB, BL Lacs etc. 
and add to them the possible contribution to extragalactic $\gamma$-rays 
from dark matter annihilations (within the framework 
of the dark matter models considered in this work). We then compare the sum total flux 
with {\it Fermi}-LAT results.

The first (Model I) of the two models are chosen based on the following 
considerations. It has 
been shown earlier, in the context of explaining the observed excess in 
Galactic Centre $\gamma$-ray signal within the energy range of around
2-8 GeV, that this excess can be well explained from the annihilation of 
WIMP (Weakly Interacting Massive Particle) dark matter of mass in the range of tens of GeV
\cite{Hooper:2010mq,PhysRevD.84.123005,BANIK2015420,Biswas:2015sva} 
which annihilates principally
to $b\bar{b}$ as primary products with the dark matter annihilation 
cross-section $\langle \sigma v \rangle$ ($v$ represents the velocity of dark matter) in 
the ball park of $10^{-26}$ cm$^3$ s$^{-1}$. The first model for particle dark matter 
considered in this word is inspired by \citet{pandeymajumdar}. 
It has been demonstrated \cite{pandeymajumdar} that the WIMP dark matter
component of this first model (describe later) satisfies these 
criteria along with the criteria for dark matter relic density given 
by Planck satellite borne experiment \cite{planck} based on the observation and analysis 
of the anisotropies of Cosmic Microwave Background Radiation (CMBR). This has also been shown in \citet{pandeymajumdar} that the WIMP component (of mass $\sim50$ GeV) in their two component dark matter model (adopted here as Model I) can well explain the observed Galactic Centre $\gamma$-ray fluxes observed by {\it Fermi}-LAT from this region. The dark mater 
candidate in the second model (Model II) 
adopted here is inspired by an established beyond Standard Model (BSM)
theory namely 
theory of universal extra dimension \cite{Cheng:2002ej}.
This has been shown 
by \cite{servant_tait,Majumdar:2003dj},
that the dark matter mass in this model should be around 900 GeV in order 
that calculated relic density for such a dark matter in this model
satisfies Planck relic density results. Being in a mass regime higher than 
that for the dark matter candidate in Model I, the $\gamma$ ray spectrum 
originated from this dark matter annihilation will have a wider energy range 
and thus raises the possibility of exploring the 
whole energy range given by the {\it Fermi}-LAT observed results for the 
upper bound of the $\gamma$-ray flux from all the 45 dwarf galaxies considered 
in this work. The same particle dark matter formalism is then adopted for the 
extragalactic $\gamma$-ray case.

The first (Model I) of the two particle dark matter models mentioned above 
is a two component dark matter 
model (obtained by minimal extension of Standard Model) 
where one component is a Feebly Interacting Massive particle 
or FIMP (we have denoted FIMP as FImP for being less massive) and the other component is a Weakly Interacting Massive 
Particle or WIMP. The model is proposed and its phenomenology 
is elaborately worked out in an earlier work 
involving two of the present authors (Ref. \cite{pandeymajumdar}). 
While the FImP component of this two component dark matter model 
could explain the phenomena such as dark matter self interactions, the 
WIMP component was useful in explaining the excess $\gamma$-rays from 
GC region when they annihilate mainly into $b\bar{b}$ to finally produce
$\gamma$-rays. The model is constructed by minimal extension of Standard 
Model with a Dirac fermion $\chi$, a real scalar $S$ and a 
pseudoscalar $\phi$. While the fermion $\chi$ 
and the scalar $S$ are singlets under SM gauge group, the fermion has an 
additional U(1)$_{\rm DM}$ charge. This prevents the fermion $\chi$ to 
interact with SM fermions ensuring stability. A Z$_2$ symmetry is imposed
on the scalar $S$. The Lagrangian is CP invariant but the CP invariance
is broken when the pseudoscalar $\phi$ acquires a vacuum expectation 
value (vev). On the other hand, 
the scalar develops a vev when the Z$_2$ symmetry is spontaneously 
broken. Thus after spontaneous breaking of the symmetries 
(SU(2)$_{\rm{L}} \times$ U(1)$_{\rm{Y}}$, Z$_2$, CP), the scalars in the theory namely 
the Higgs $H$, $S$ and $\phi$ acquire vev and their real components 
mix together. Three mass eigen states $h_1$, $h_2$ and $h_3$ are obtained (small mixing angles, $\theta_{12} \sim 10^{-2}$, $\theta_{13}\sim 10^{-13}$ and $\theta_{23} \sim 10^{-15}$) after diagonalisation of mass matrix while eigen state $h_1$ is identified with the physical Higgs with mass 125.5 GeV, $h_2$ is identified with the pseudo scalar considered in the model. Since $h_1$ is identified with SM Higgs, one requires to consider the collider bounds for the limit on $R_1$ (SM Higgs signal strength) which is expressed as the ratio of total decay width of the SM Higgs and the calculated total decay width of $h_1$. One also requires to consider invisible decay branching ratio ($Br_{inv}^1$) of SM like Higgs $h_1$ and this parameter is fixed using collider constrains. The lightest mass eigenstate after diagonalisation (with 
small mixing with other scalars) is taken to be the FImP candidate. But 
in this work, in order to calculate the $\gamma$-rays from the annihilation 
of dark matter in each of the chosen 45 dwarf galaxies (as also the extragalactic 
$\gamma$-ray case), the WIMP component 
which is the Dirac singlet fermion (in this WIMP-FImP model) is useful. The 
WIMP candidate in this model interacts with SM sector through Higgs 
portal. In Ref.~\citet{pandeymajumdar}, this has been shown that 
the excess $\gamma$-rays from the Galactic Centre within the energy range 
2 GeV - 8 GeV as reported by {\it Fermi}-LAT can be well explained 
by the WIMP component (fermion $\chi$) of this WIMP-FImP model if 
the WIMPs ($\chi$) self annihilate to $b\bar{b}$ which in turn 
produces secondary $\gamma$-rays. The cross-section 
for the channel $\chi \chi \rightarrow b\bar{b}$ is calculated and 
computed in Ref. \citet{pandeymajumdar}  for certain model  benchmark points. 
The computations of annihilation cross-section of the process $\chi \chi \rightarrow b \bar{b}$ require the coupling and other factors (the expression for this cross-section is given in Appendix of \citet{pandeymajumdar}) to be determined by constraining the model interaction lagrangian with theoretical bounds as well as experimental bounds and collider bounds. The process $\chi \chi \rightarrow b \bar{b}$ is mediated by the scalar $h_1$ and $h_2$ and one needs the coupling `$g$' of the pseudo scalar ($h_2$) with the dark matter fermion $\chi$ as also the other couplings of $\chi$ with $h_1$. The coupling $g$ is a parameter in this model (the expression of the couplings of the fermions with Higgs are known). Again since Model I is two component dark matter model (the WIMP component of which is considered here (similar to what is considered in \citet{pandeymajumdar})), the fraction of WIMP component $\chi$ that contributes to generate required Galactic Centre $\gamma$-ray excess in \citet{pandeymajumdar} is denoted by $f_{\chi}$. The annihilation cross-section $\langle \sigma v \rangle_{\chi \chi \rightarrow b \bar{b}}$ will also be weighted by this fraction $f_{\chi}$. Thus $f_{\chi}$ is also a parameter of the model. The allowed ranges of all these parameters are obtained by constraining the model interaction lagrangian with theoretical (unitarity, perturbativity etc.) as well as experimental constraints (e.g. PLANCK relic density results) and collider constraints. In Ref.~\citet{pandeymajumdar} the several benchmark values of model parameters are given for computing the Galactic Centre $\gamma$-ray excess and comparing with the {\it Fermi}-LAT results for the same. In the present work, we adopt one of such benchmark point and compute the $\gamma$-ray fluxes for all the 45 dSphs by similar consideration of annihilation of the WIMP component $\chi$ (of the two component model discussed above) to $b\bar{b}$ and compare them with upper bounds of all those 45 dSphs given by {\it Fermi}-LAT experiment \cite{fermilat,Fermi-LAT:2016uux}. This benchmark point (set of values of the parameters) are shown in Table~\ref{tab:model1}. 
The astrophysical  
$\mathcal{J}$-factor values required to compute the fluxes are
obtained from different observational groups for the 
dwarf galaxies.

The other particle dark matter candidate considered in this work is from 
a BSM theory and this candidate 
is Kaluza-Klein (KK) dark matter (Model II) inspired 
by the theories of extra dimensions 
\cite{Cheng:2002ej,servant_tait,Hooper:2007gi,Majumdar:2003dj}. If only one 
spatial extra dimension 
is considered and this extra dimension is compactified over a circle of 
compactification radius $R$, say, then the effective four dimensional theory
as obtained by integrating the extra spatial dimension over the periodic 
coordinate (($y \rightarrow y + 2\pi R$), compactification over a circle), 
gives rise to a tower of Kaluza-Klein modes with mass of each mode given
by $m_k = k/R$, where $k$ is called the Kaluza-Klein number or KK number.
As KK number is associated with the quantized momentum in compactified 
dimension ($E^2 = $ {\bf p}$^2 + m_k^2$), the KK number is conserved 
and hence the Lightest Kaluza-Klein particle or LKP is stable and can be 
a candidate for dark matter. 

In this work, we consider a KK dark matter candidate in an extra 
dimensional model namely Universal Extra Dimensional model 
(UED) \cite{Cheng:2002ej,servant_tait,ued2,ued3}. In this model, each of the SM 
field can propagate in the extra dimension and every SM particle 
has a KK tower. 
But since the SM fermions are chiral, in order to obtain chiral KK 
counterpart of the SM fermions in UED model, the compactification of 
the extra dimension is to be made over a $S^1/\rm{Z}_2$ orbifold 
(instead of compactifying just over a circle $S^1$ with compactification 
radius $R$) where a reflection symmetry $\rm{Z}_2$ is imposed under which 
the extra coordinate $y\rightarrow -y$ and the fields are even or odd. 
Thus the chirality of a fermion can be identified in the extra dimension. 
The orbifold has now two boundary points at 0 and $\pi R$. But this breaks 
translational symmetry in the $y$ direction and the KK momentum 
is no more conserved. Therefore the KK number (k) is also not conserved and the LKP is no more stable. But, for the transformation 
$y \rightarrow y+\pi R$, the KK modes remain invariant for even KK 
number but odd KK modes change sign. Thus we have a quantity 
called $(-1)^{KK}$ - the KK parity - which is a good symmetry for this 
transformation and hence conserved. The conservation of KK parity ensures 
LKP in UED model 
is stable. In the present work, the LKP dark matter candidate 
in UED model is the first KK partner $B^1$ of the hypercharge gauge boson.

We have taken a range of masses for the chosen  
KK dark matter candidate $B^1$ and demonstrate how well the 
$\gamma$-rays produced from 
the annihilation of such a dark matter candidate 
agrees with the observational results for all the dwarf galaxies considered. 
The range of masses for these KK particles are so chosen that the PLANCK
limits for the dark matter relic densities are satisfied. For continuum 
$\gamma$ signal from $B^1 B^1$ annihilation one needs to consider 
the channel $B^1 B^1 \longrightarrow qq$ ($q$ denotes the quarks). 
The annihilation cross-sections ($\langle \sigma_{qq} v \rangle $) 
for this channel are calculated following \cite{Cheng:2002ej}. It is to 
be noted that the interaction coupling for the process $B^1 B^1\rightarrow qq$ 
is computable for a given mass of the dark matter candidate $B^1$ \cite{Cheng:2002ej}. 
The only parameter here is the mass of $q^1$ which is the first KK partner of 
quark $q$ in the UED model. The parameter is rewritten as $r=\dfrac{m_{q^1}-m_{B^1}}{m_{B^1}}$, 
where, $m_{q^1}$, $m_{B^1}$ are the masses of $q^1$ and $B^1$ respectively. In 
this work, we have varied parameter $r$ in such a way that $m_{B^1}$ is in the allowed mass range 
(discussed earlier) and $m_{q^1}>m_{B^1}$ is maintained. We do not find any significant changes in the result.

We then extend our analyses for extragalactic $\gamma$-rays also. The observed 
extragalactic $\gamma$-ray signal may contain the component of $\gamma$-ray 
from dark matter annihilations at extragalactic sources (\cite 
{Ullio:2002pj, Bergstrom:2001jj, Gao:1991rz, Stecker:1978du, Taylor:2002zd, 
Ng:2013xha}. The extragalactic $\gamma$-rays can have many components
other than those possibly from dark matter annihilations. There are attempts
to extract dark matter annihilation signals from the extragalactic 
$\gamma$-ray background or EGB \cite{Calore:2013yia, Cholis:2013ena, 
Tavakoli:2013zva, Sefusatti:2014vha, Ajello:2015mfa,DiMauro:2015tfa, 
DiMauro:2015ika, Ackermann:2015tah}. The possible 
contribution to the EGB may come from BL Lac objects, millisecond 
pulsars, radio galaxies etc. More detailed knowledge and their possible 
contribution to the EGB not only helps to look for any such dark matter 
annihilation signals beyond the EGB  but also is useful to put 
stringent bound on dark matter annihilation cross-sections. With both the 
particle dark matter models considered here, we have made an attempt 
in this work to estimate whether any significant signal from the 
dark matter annihilation can be obtained from 
extragalactic sources. For the extragalactic also, the same benchmark points 
given in Table~\ref{tab:model1} and Table~\ref{tab:model} are used.

In an earlier work (\cite{Modak:2015uda}),  
similar analyses have been performed. But in that analyses only 18 dSphs 
were considered with inert doublet dark matter but in the preset analyses 
we take into account as many as 45 dSphs. Also, in this work, 
two dark matter models are considered under which one dark matter 
candidate is the Higgs portal 
component of a two component DM (\cite{pandeymajumdar}) while the other 
is a particle dark matter inspired by theories of extra dimensions 
(\cite{Cheng:2002ej,servant_tait,Hooper:2007gi,Majumdar:2003dj}).

The paper is organised as follows. In Sect.~\ref{sec:flux} we give the 
formalism to 
calculate $\gamma$-ray flux. Sect.~\ref{sec:d_gal} deals with the 
observational data, the 
calculations and results for the dwarf galaxies and the comparison of the 
computed results with observational bounds. The calculational 
procedures for the estimation of extra galactic $\gamma$-ray background 
and the contribution from possible dark matter annihilation are given 
in Sect.~\ref{sec:ex_gal}. Finally in Sect.~\ref{summ} we conclude 
with a summary and some remarks.

\section{Formalism for $\gamma$-ray flux calculations in case of dwarf galaxies from dark matter annihilation} \label{sec:flux}

{\begin{table}
	\centering
	\caption{The model parameter considered for the calculation on $\gamma$-ray fluxes in Model I. $v_1$ is the vev of SM Higgs.\label{tab:model1}}
	\begin{tabular}{ccccccc}
		\hline
		$M_{\chi}$ & $v_1$ & $g$ & $R_1$ & $Br^1_{inv}$ & $f_{\chi}$ & $f^2_{\chi} \langle \sigma v \rangle_{b \bar{b}} $\\
		GeV & GeV & & & & & $10^{-26}\rm{cm}^3 \rm{s}^{-1}$\\
		\hline
		50 & 246 & 0.11 & 0.99 & 0.021 & 0.89 & 1.62\\
		\hline
	\end{tabular}
\end{table}

The observed flux from cosmic dark matter source depends significantly 
on the dark matter annihilation cross-section 
$\langle\sigma v\rangle$ \cite{fermilat} as well as the total DM contained 
within the solid angle subtended by the source at the observer 
(the astrophysical $\mathcal{J}$-factor). Analytically the 
$\mathcal{J}$-factor can be calculated as
\begin{equation}
\mathcal{J} = \int_{l.o.s}\rho(r)^2ds=r_{\odot} \rho_{\odot}^2 J.
\label{eq_jfactor}
\end{equation}
In the above, $\rho_{\odot}$ ($0.3~\rm{GeV/cm^3}$) is the 
dark matter density 
at the distance $r_{\odot}$ ($8.33~\rm{kpc}$) from the Galactic 
Centre (at the solar system). In the above equation, $J$ represents the dimensionless form 
of $\mathcal{J}$-factor given by,
\begin{equation}
J = \int_{l.o.s}\frac{1}{r_\odot} \left(\frac{\rho(r)}
{\rho_{\odot}}\right)^2 ds,
\label{eq_jfactor_dl}
\end{equation}
where $\rho(r)$ is the DM density at radial distance $r$ from the 
Galactic Centre and $\rho(r)$ in a dark matter halo can be parametrised as 
$\rho(r) = \rho_s g(r/r_s)$, where $\rho_s$ is a scale density
and $g(r/r_s)$ gives the nature of density function with $r$ and 
$r_s$ is a characteristic scale distance. In this case $r_s = 20$~kpc for dwarf galaxy calculations. The radial distance $r$ can be expressed in terms of 
the line of sight $s$ as,
\begin{equation}
r=\begin{array}{ll}
\sqrt{s^2+r_{\odot}^2-2sr_{\odot}\cos{l}\cos{b}}
&~~~~~~~~l,~b ~\rm{coordinate},\\
\sqrt{s^2+r_{\odot}^2-2sr_{\odot}\cos{\theta}}
&~~~~~~~~r,~\theta ~\rm{coordinate}.\\
\end{array}
\label{eq_r}
\end{equation}  
We adopt three density profiles for computation of $\rho(r)$ and 
those three profiles are given in 
Table~\ref{app:haloprofile}. 
\begin{table*}
	\centering
	\caption{Dark matter halo profiles \label{app:haloprofile}}
	\begin{tabular}{|llcl|}
		\hline
		NFW \cite{Navarro:1995iw,Navarro:1996gj} & ~~$\rho_{\rm{NFW}}$ &$=$& $\rho_s\frac{r_s}{r}\left(1+\frac{r}{r_s}\right)^{-2}$\\
		Einasto \cite{einasto} & ~~$\rho_{\rm{Ein}}$ &$=$& $\rho_s\exp{\left[-\frac{2}{\alpha}\left\{\left(\frac{r}{r_s}\right)^{\alpha}-1\right\}\right]}$\\
		Burkert \cite{burkert1,burkert2} & ~~$\rho_{\rm{Bur}}$ &$=$& $\frac{\rho_s}{\left(1+r/r_s\right)\left(1+\left(r/r_s\right)^2\right)}$\\
		\hline
	\end{tabular}
\end{table*}

The differential $\gamma$-ray 
flux due to dark matter annihilation of mass $M_{\chi}$ 
is given by \cite{cirelli},
\begin{equation}
\frac{d\phi}{d\Omega dE_{\gamma}}=\frac{1}{8 \pi \alpha} 
\sum_{f} \frac{\langle\sigma v\rangle_f}{M_{\chi}^2} 
\frac{dN_{\gamma}^f}{dE_{\gamma}}\mathcal{J},
\label{eq_flx}
\end{equation} 
where $\alpha=1$ and $f$ indicates the 
final state particle. 

As mentioned earlier, we have considered two particle dark matter models 
in this work.
One is a two component WIMP-FImP model where the WIMP component $\chi$ 
of mass around 50 GeV (Table~\ref{tab:model1}) contributes 
to $\gamma$-rays (Model I) by their self 
annihilation (via a Higgs portal) leading primarily to $b\bar{b}$. 
The expression of the cross-section 
for the process $\chi \chi \longrightarrow b\bar{b}$ is given in the 
Appendix of \cite{pandeymajumdar}. The annihilation cross-section is computed to be $\langle \sigma v\rangle=1.62\times 10^{-26} \rm{cm^3sec^{-1}}$. 
The computations of annihilation cross-section require numerical values of the various couplings involved in different interactions. These couplings are the particle dark matter model parameters and are discussed in \citet{pandeymajumdar}. In Table~\ref{tab:model1} we furnish a set of benchmark values taken from the allowed region of these parameters discussed in \citet{pandeymajumdar}. The set of values of the couplings is used in present work.
It is to be noted that, these coupling parameters are in agreement with the 
bounds and the constraints coming from the theoretical considerations such as 
vacuum stability conditions, perturbativity and unitarity conditions etc. and they 
also satisfy the various bounds given by the experimental observations, e.g., 
the collider physics bounds (LHC bounds), the PLANCK observational results 
for the dark matter relic density and the upper limit on the dark matter - 
nucleon scattering cross-section obtained from different dark matter direct 
detection experiments. All the above mentioned constraints and bounds imposed 
on the model parameters (such as various interaction couplings etc.) which 
are necessary in order to compute the annihilation cross-section for the 
channel $\chi \chi \rightarrow b \bar{b}$, have been elaborately discussed in 
\cite{pandeymajumdar}. The parameters in 
Table~\ref{tab:model1} are within the allowed range of model parameter space and they also 
respect all the necessary constraints and bounds. Same set of coupling parameter value and the same formalism are adopted for extragalactic case also.

The other is a KK dark matter ($B^1$) in an extra dimensional model 
(Model II) having a mass of about 900 GeV 
(Table~\ref{tab:model}) 
which self annihilates to the primary product $qq$ and yields 
$\gamma$-rays as the end product. 
For the case of $B^1$ dark matter, the annihilation cross-section 
$B^1 B^1 \longrightarrow qq$ is computed from the expression 
\cite{Cheng:2002ej} 
\begin{equation}
\langle \sigma_{qq} v \rangle = \frac {q^4} {9\pi \cos^4\theta_W}
\left [ \frac {Y_{q_L^1}^4} {m_{B^1}^2 + m_{q_L^1}^2} + L \rightarrow R 
\right ]
\end{equation}
where $q_L^1$ is the first KK partner of the quark $q_L$, $Y_{q_L^1}$
and $m_{q_L^1}$ are respectively 
the corresponding hypercharge and mass while $\theta_W$ is the Weinberg 
angle. The mass $m_{q_L^1}$ are fixed by defining a parameter 
$r = (m_{q^1} - m_{B^1})/m_{B^1}$ (\cite{Cheng:2002ej}) and then adopting 
a suitable value for $r$. It is also to be noted that the LKP dark matter
candidate $B^1$ in this case is the first KK partner of the hypercharge gauge
boson. It is seen that the mass of this dark matter candidate should be $\sim$900 GeV 
for its relic density to satisfy the PLANCK result \cite{planck}.
In the limit in which electroweak symmetry breaking (EWSB) is neglected,
there will be no channels with vector gauge bosons as primary products 
and only 2\% of the 
annihilation goes into Higgs \cite{servant_tait}. Moreover 
\cite{Cheng:2002ej} shows that the channel $B^1 B^1 \rightarrow e^+e^-$
yields narrow peaks for positrons and for computation of continuum photon 
signal the relevant annihilation cross-section is $\langle \sigma_{qq} v 
\rangle$. The choice of dark matter masses in both the models are shown in 
Table~\ref{tab:model}. Same formalism of dark matter annihilation with same dark matter mass is adopted for the extragalactic $\gamma$-ray flux calculations.
 
The $\gamma$ fluxes are calculated (for the chosen 
dark matter candidates) by computing the ${\cal J}$ factor with each of the 
three density profiles of Table~\ref{app:haloprofile}. 
The fluxes are 
also computed  with the ${\cal J}$ factors estimated and published 
by other groups \cite{jfact1,jfact2}. These density profiles are plotted in Fig.~\ref{fig:profiles}.
\begin{table}
	\centering
	\caption{The masses of the dark matter in both the models adopted in the work.\label{tab:model}}
	\begin{tabular}{lc}
		\hline
		Model & $M_{\chi}$ in GeV\\
		\hline
		Model I (\cite{pandeymajumdar}) &$50$\\
		Model II (\cite{Cheng:2002ej,servant_tait}) &$900$\\
		\hline
	\end{tabular}
\end{table}

\section{The $\gamma$-ray flux calculations for the dwarf galaxies and their comparison with the observations}\label{sec:d_gal} 
DM rich dwarf spheroidal galaxies (dSphs) have turned out to be important cosmological sites 
to probe and understand the nature of dark matter and its astrophysical 
implications. 
The satellite observations by {\it Fermi}-LAT \cite{fermilatold} 
as well as later Dark Energy Survey (DES) and {\it Fermi}-LAT 
collaboration 
reveal a sum of $45$ dwarf spheroidal galaxies in the energy 
range $0.5\sim500$ GeV \cite{fermilat}. The details of these 
dSphs are furnished in Table~\ref{tab:dsphs}. 
In Table~\ref{tab:dsphs} the ${\cal J}$ factors and their uncertainties
are also given for those 45 dSphs. These {\it Fermi}-LAT observations upper 
bounds are obtained from \cite{fermilat,Fermi-LAT:2016uux}. The upper bounds of 
$\gamma$-ray fluxes from those dSphs 
are given in Fig.~\ref{fig:grd_1}. 

In the present work we have estimated $\gamma$-flux for all of 
those $45$ dSphs tabulated in Table~\ref{tab:dsphs} assuming 
that the dark matter in those dSphs annihilate to produce $\gamma$.
The computations for $\gamma$ flux for each of the two DM candidates have been performed following Eqs. 
(1) - (5). The DM candidates and models as also the chosen DM masses are
already discussed and in      
Table~\ref{tab:model} the benchmark mass points are given. Note that the cross-section given in Eq.~5 is for KK dark matter only (Model II).

Figs.~\ref{fig:grd_1} and \ref{fig:grd_2} show upper bound of 
$\gamma$-ray 
flux given by the {\it Fermi}-LAT observation as well as the computed $\gamma$-ray flux for the two particle dark 
matter candidates. 
In Figs.~\ref{fig:grd_1} and \ref{fig:grd_2} the upper bound 
of the $\gamma$-ray flux 
for all the 45 galaxies (as given by {\it Fermi}-LAT collaboration) are 
shown with green arrows pointing downwards. 
Integrated $\mathcal{J}$-factor over a solid angle 
of $\Delta\Omega=2.4\times10^{-4}$ sr (field of view of 
{\it Fermi}-LAT $\sim 0.5^o$) are measured from stellar kinematics data. 
The numerical values of $\mathcal{J}$-factor for all dSphs (obtained from observational data) are 
tabulated in Table~\ref{tab:dsphs}. 

The computations of flux for the DM candidates in both the models 
are made as follows. Integrated $\mathcal{J}$-factor over a solid angle 
of $\Delta\Omega=2.4\times10^{-4}$ sr (field of view of 
{\it Fermi}-LAT $\sim 0.5^o$) are measured from stellar kinematics data. 
The numerical values of $\mathcal{J}$-factor for all dSphs (obtained from observational data) as well as the flux estimations for the case of both the models I and II using the $\cal{J}$-factors are 
tabulated in Table~\ref{tab:dsphs}. 
The flux estimations for the case of both the 
Models I and II are first made using the ${\cal J}$ factors given in 
Table~\ref{tab:dsphs}. 
The spread of each of these calculated fluxes due 
to the uncertainties of the ${\cal J}$ factors (given in Table~\ref{tab:dsphs})
are also calculated in case of
each of the two dark matter models considered. The fluxes and their 
spreads thus 
estimated with both the DM candidates for all the 45 dSphs are shown in
Figs.~\ref{fig:grd_1} and \ref{fig:grd_2}. The $\gamma$ flux for DM in 
Model I (WIMP component $\chi$ with mass $\sim$50 GeV of two component 
WIMP-FImP model) are shown by a 
black line (for the central values of ${\cal J}$ in Table~\ref{tab:dsphs})
and the estimated spread of these computed fluxes due to uncertainties 
in corresponding ${\cal J}$ values are shown by yellow bands  
in each of the 45 plots (for 45 dwarf galaxies) spreaded over 
Figs.~\ref{fig:grd_1} and \ref{fig:grd_2}. Similar
estimations of the $\gamma$ flux and their uncertainties for the DM candidate
$B^1$ (KK dark matter from extra dimensional model with mass $\sim$900 GeV; Model II)
are shown with blue central lines with uncertainty spreads shown in pink in 
each of the 45 plots (of Figs.~\ref{fig:grd_1} and \ref{fig:grd_2}). 

The fluxes are also estimated for both the dark matter candidates 
(Model I and Model II)
following Eqs.~\ref{eq_jfactor}-\ref{eq_flx} by explicitly computing
the ${\cal J}$ with each of the three dark matter density profiles given 
in Table~\ref{app:haloprofile}. These are NFW, Einasto and Burkert 
DM density profiles. They are shown in Figs.~\ref{fig:grd_1} and 
\ref{fig:grd_2} in red, black and blue dashed lines respectively 
for the WIMP dark matter in Model I and in red, black and blue dotted lines for the KK dark matter of  Model II.
It can be seen that for most of the cases, the results using NFW 
profiles almost coincide with those using Einasto profile while 
distinction can be made for the flux results with Burkert profile.
This may be understood from the natures of the profiles 
(Fig.~\ref{fig:profiles}). While both 
NFW and Einasto profiles are cuspy in nature, the Burkert profile is
flat and isothermal in nature. Also Burkert profile has been used earlier
to analyse the dwarf galaxy rotation curves (\cite{mnras1,burkert1}).

\begin{figure}
	\centering
	\includegraphics[width=0.5\textwidth] {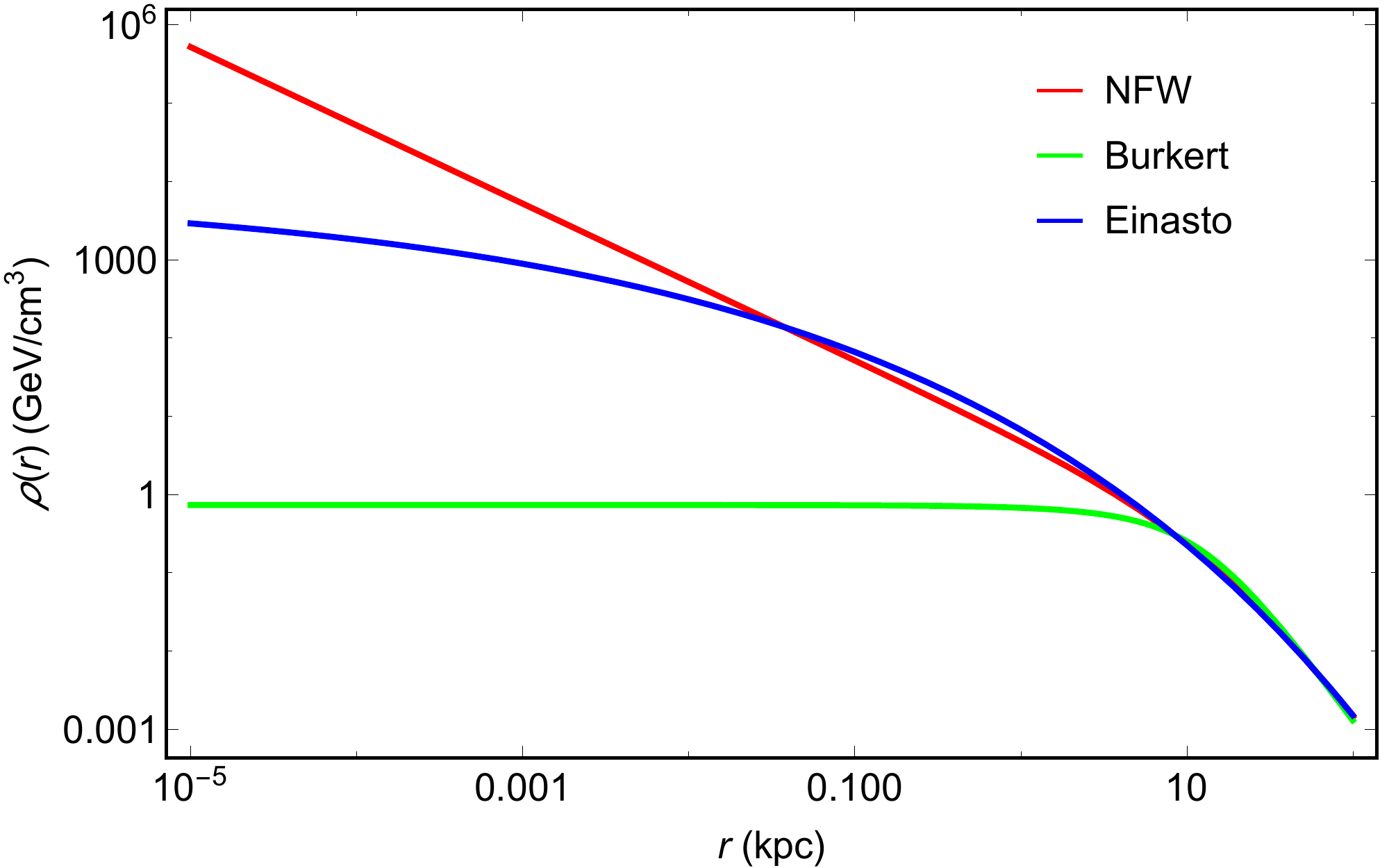}
	\caption{Galactic dark matter halo density profiles.\label{fig:profiles}}
\end{figure} 

It appears from Figs.~\ref{fig:grd_1} and \ref{fig:grd_2} 
that for Model I (the WIMP component of a WIMP-FImP model) 
with all the three density profiles, the fluxes are below the observational
upper limits of all the 45 dSphs. Similar results for the KK dark matter in
Model II (the Kaluza-Klein model) are shown in Figs.~\ref{fig:grd_1} and 
\ref{fig:grd_2} with colour codes mentioned above. It can be seen that KK dark matter 
also respects the observational upper bounds of all the dSphs $\gamma$-fluxes considered here.
Moreover, it can also be seen form Figs.~\ref{fig:grd_1} and 
\ref{fig:grd_2} that wider range (in comparison to what obtained in case of Model I) of $\gamma$-ray flux can be achieved when KK dark matter (Model II) is considered.

\begin{figure*}
	\centering
	\includegraphics[width=0.8\textwidth] {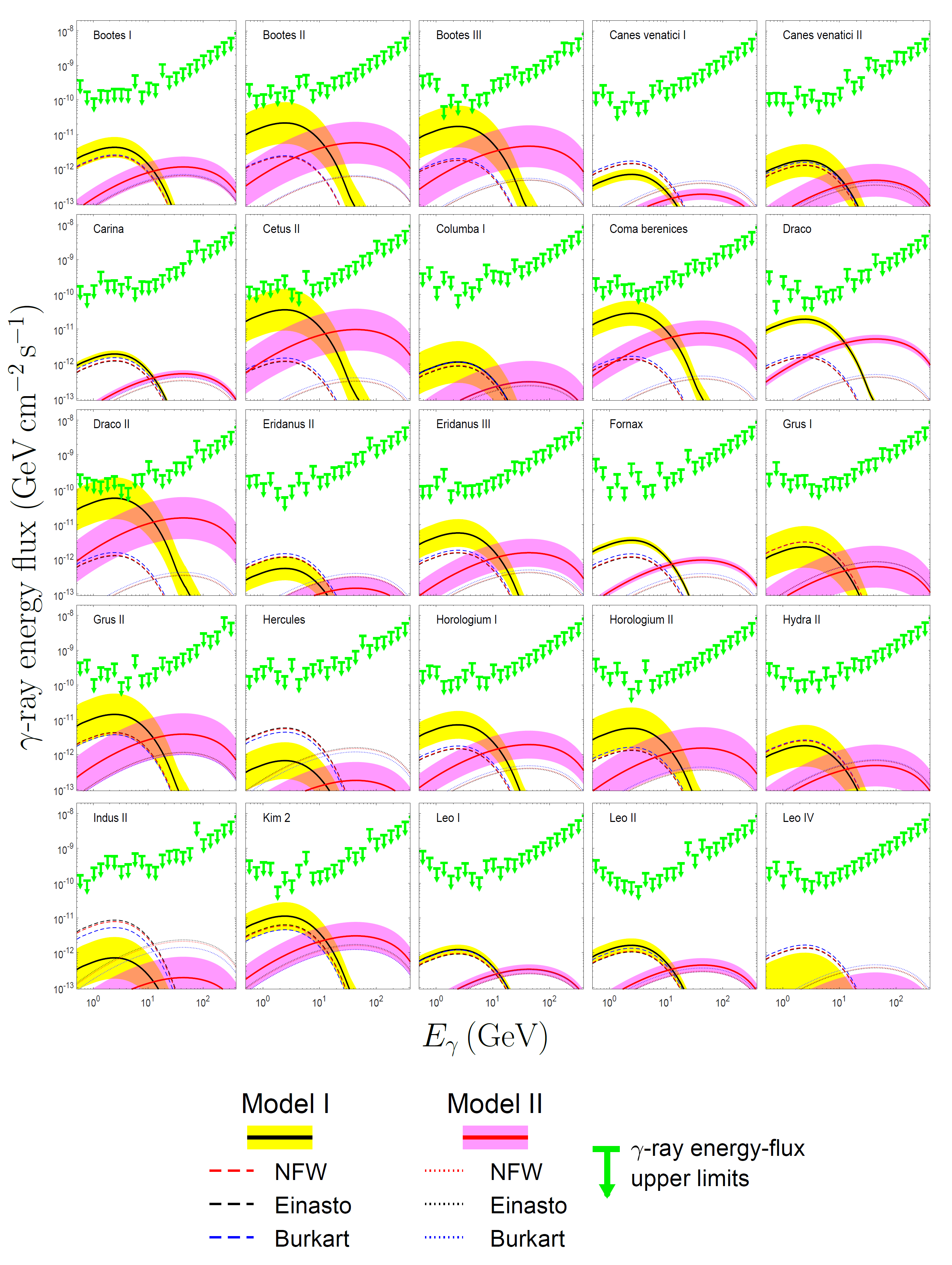}
	\caption{$\gamma$-ray fluxes from dark matter anihilations for each of the 
		dark matter candidates in Model I (50 GeV fermionic WIMP) and  
		Model II (900 GeV Kaluza-Klein dark matter) calculated for each of 
		the 25 dwarf galaxies and their comparisons with experimental upper bounds 
		of $\gamma$-ray flux (shown by green coloured downward arrows) 
		for each of the dwarf spheroidals. The flux calculations
		with the ${\cal J}$ factors from Table~\ref{tab:dsphs} and its uncertainty spreads 
		are shown by black solid line and yellow band respectively when 
		Model I is considered and the same for the DM candidate of Model II 
		are shown by pink solid line and pink band respectively. The ${\cal J}$ factors for both Model I and Model II The computed using three dark matter density profiles which are shown  with dashed lines and dotted lines of different colours for comparisons. See text for details.\label{fig:grd_1}}
\end{figure*}

\begin{figure*}
	\centering
	\includegraphics[width=0.8\textwidth] {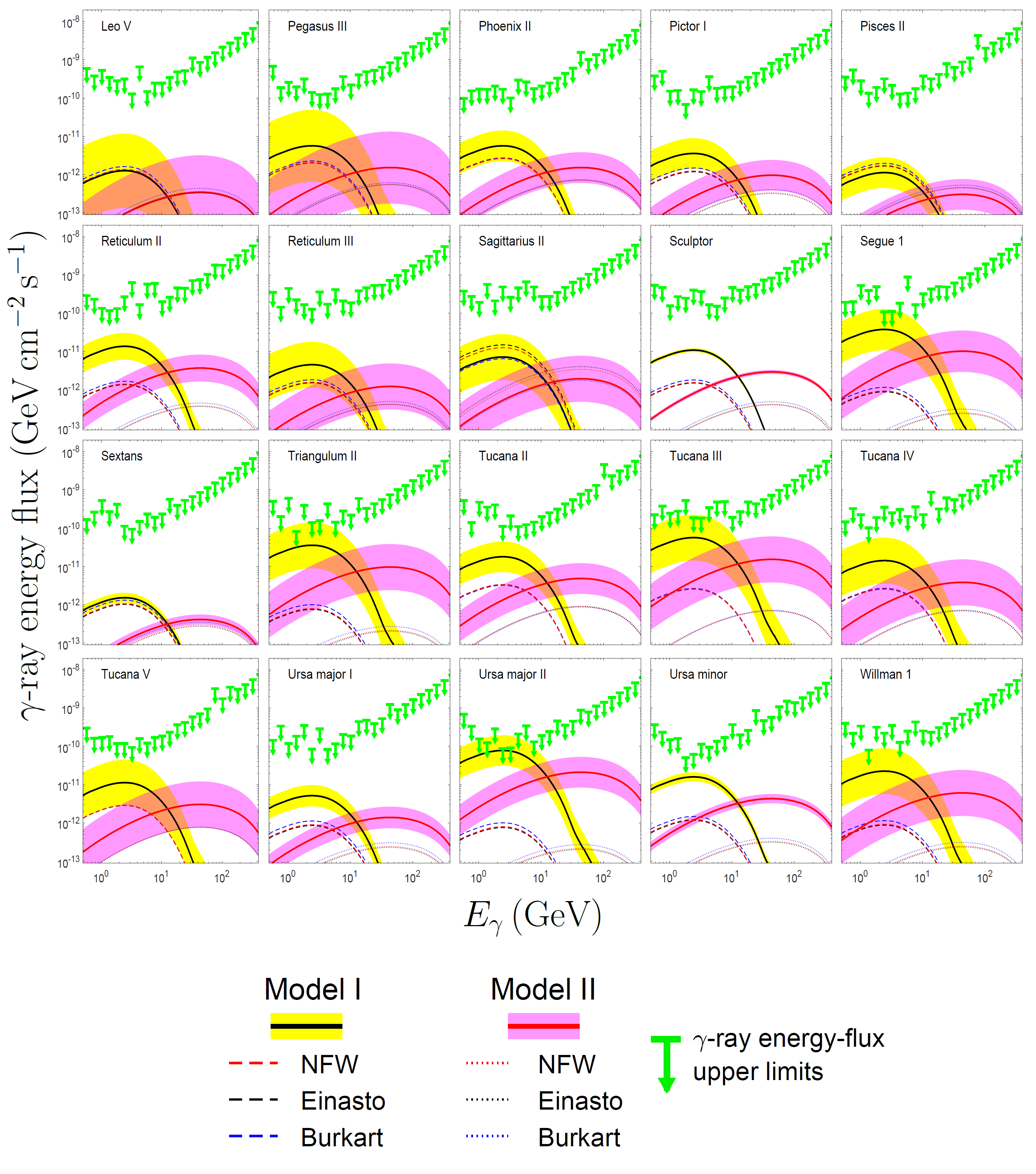}
	\caption{Same as Fig.\ref{fig:grd_1} but for the rest 20 dwarf galaxies. See text for details.\label{fig:grd_2}}
\end{figure*}
\begin{table*}
	\centering
	\caption{Latitude, longitude, distance and $\mathcal{J}$-factor for individual dSphs \cite{jfact1,jfact2}.\label{tab:dsphs}}
	\begin{tabular}{lcccc}
		\hline
		dSphs name & Longitude & 
		Latitude & Distance & 
		$\log_{10}\mathcal{J}$\\
		& $l$ (deg) & $b$ (deg) & 
		(kpc) & ($\log_{10}\left[\rm{GeV^2 cm^{-5} sr}\right]$)\\
		\hline
		Bootes I &$358.1$&$69.6$&$66$&$18.17\pm0.30$\\
		Bootes II &$353.7$&$68.9$&$42$&$18.90\pm0.60$\\
		Bootes III &$35.4$&$75.4$&$47$&$18.80\pm0.60$\\
		Canes Venatici I &$74.3$&$79.8$&$218$&$17.42\pm0.16$\\
		Canes Venatici II &$113.6$&$82.7$&$160$&$17.82\pm0.47$\\
		Carina &$260.1$&$-22.2$&$105$&$17.83\pm0.10$\\
		Cetus II &$156.47$&$-78.53$&$30$&$19.10\pm0.60$\\
		Columba I &$231.62$&$-28.88$&$182$&$17.60\pm0.60$\\
		Coma Berenices &$241.9$&$83.6$&$44$&$19.00\pm0.36$\\
		Draco &$86.4$&$34.7$&$76$&$18.83\pm0.12$\\
		Draco II &$98.29$&$42.88$&$24$&$19.30\pm0.60$\\
		Eridanus II &$249.78$&$-51.65$&$330$&$17.28\pm0.34$\\
		Eridanus III &$274.95$&$-59.6$&$95$&$18.30\pm0.40$\\
		Fornax &$237.1$&$-65.7$&$147$&$18.09\pm0.10$\\
		Grus I &$338.68$&$-58.25$&$120$&$17.90\pm0.60$\\
		Grus II &$351.14$&$-51.94$&$53$&$18.70\pm0.60$\\
		Hercules &$28.7$&$36.9$&$132$&$17.37\pm0.53$\\
		Horologium I &$271.38$&$-54.74$&$87$&$18.40\pm0.40$\\
		Horologium II &$262.48$&$-54.14$&$78$&$18.30\pm0.60$\\
		Hydra II &$295.62$&$30.46$&$134$&$17.80\pm0.60$\\
		Indus II &$354$&$-37.4$&$214$&$17.40\pm0.60$\\
		Kim 2 &$347.2$&$-42.1$&$69$&$18.60\pm0.40$\\
		Leo I &$226$&$49.1$&$254$&$17.64\pm0.14$\\
		Leo II &$220.2$&$67.2$&$233$&$17.76\pm0.2$\\
		Leo IV &$265.4$&$56.5$&$154$&$16.40\pm1.15$\\
		Leo V &$261.86$&$58.54$&$178$&$17.65\pm0.97$\\
		Pegasus III &$69.85$&$-41.81$&$205$&$18.30\pm0.94$\\
		Phoenix II &$323.69$&$-59.74$&$95$&$18.30\pm0.40$\\
		Pictor I &$257.29$&$-40.64$&$126$&$18.10\pm0.40$\\
		Pisces II &$79.21$&$-47.11$&$182$&$17.60\pm0.40$\\
		Reticulum II &$266.3$&$-49.74$&$32$&$18.68\pm0.35$\\
		Reticulum III &$273.88$&$-45.65$&$92$&$18.20\pm0.60$\\
		Sagittarius II &$18.94$&$-22.9$&$67$&$18.40\pm0.60$\\
		Sculptor &$287.5$&$-83.2$&$86$&$18.58\pm0.05$\\
		Segue 1 &$220.5$&$50.4$&$23$&$19.12\pm0.54$\\
		Sextans &$243.5$&$42.3$&$86$&$17.73\pm0.13$\\
		Triangulum II &$140.9$&$-23.82$&$30$&$19.10\pm0.60$\\
		Tucana II &$328.04$&$-52.35$&$58$&$18.80\pm0.40$\\
		Tucana III &$315.38$&$-56.18$&$25$&$19.30\pm0.60$\\
		Tucana IV &$313.29$&$-55.29$&$48$&$18.70\pm0.60$\\
		Tucana V &$316.31$&$-51.89$&$55$&$18.60\pm0.60$\\
		Ursa Major I &$159.4$&$54.4$&$97$&$18.26\pm0.28$\\
		Ursa Major II &$152.5$&$37.4$&$32$&$19.44\pm0.40$\\
		Ursa Minor &$105$&$44.8$&$76$&$18.75\pm0.12$\\
		Willman 1 &$158.6$&$56.8$&$38$&$18.90\pm0.60$\\
		\hline
	\end{tabular}
\end{table*}

\section{Extragalactic $\gamma$-ray background and extragalactic $\gamma$-rays from dark matter annihilations} \label{sec:ex_gal}

In this section, we compute the defused extragalactic $\gamma$-ray flux from dark matter annihilation and compared with different possible backgrounds. 
Here we like to mention that, the saying models (Model I and Model II as described earlier) are adopted for particle dark matter candidate and the same set of model parameter values  given in Table~\ref{tab:model1} and Table~\ref{tab:model} re used for computing the dark matter annihilation cross-section in extragalactic case also.
The $\gamma$-ray flux from dark matter annihilation could have extragalactic origins too and probing 
such $\gamma$-rays could be effective not only for indirect detection of extragalactic DM but also 
to understand their origins \cite{Stecker:1978du,Taylor:2002zd,Gao:1991rz,Ando_2005,Bergstrom:2001jj,Oda_2005,Ullio:2002pj,Ng:2013xha,Pieri_2008}. But whether such $\gamma$-ray signals can be identified 
by terrestrial telescopes depend on the background $\gamma$-rays from different other types of 
extragalactic sources. Therefore, to study the $\gamma$-rays from extragalactic
dark matter annihilation, one needs to estimate the flux from other 
possible sources that can contribute to the backgrounds for such observations.

In order to explore the possibilities that $\gamma$-ray signals from 
the extragalactic dark matter annihilations (indirect DM signals) could 
be detected with significance, we also compute the $\gamma$-ray signals
from other possible non-DM origins of extragalactic $\gamma$-rays. Such 
non-DM origins include BL Lac objects, quasars, pulsars, 
Gamma Ray Bursts (GRB) etc. 
For many of these sources, the natures of spectra are 
found to follow roughly a power law.  
A list of such sources and the corresponding $\gamma$-ray flux 
(power law or other forms) from these sources are furnished
later in Table~\ref{tab:power_sp}.

The satellite borne experiment namely {\it Fermi}-LAT furnished
their observed results for extragalactic $\gamma$-ray flux. In this 
section, we compute the sum of the $\gamma$-rays from 
extragalactic DM annihilations (for each of the DM candidate
in Model I and Model II) and from other possible non-DM sources. 
We then compare our results with those observed by {\it Fermi}-LAT \cite{2010PhRvL.104j1101A,Ackermann:2014usa}.

The rate of photons emitted from volume element $dV$ from the sky depends on several factors mainly the halo mass function $dn/dM$ as a function of mass $M$ and redshift $z$, the differential photon energy spectrum $\frac{d\mathcal{N}_{\gamma}}{dE}(E,M,z)$, the attenuation factor ($e^{-\tau}$) of the extragalactic $\gamma$-rays etc. The rate of photons emitted from volume element $dV$ having energy 
ranges $E+dE$ and observed by detector having effective area $dA$ (with time interval $dt$ and redshifted energy interval $dE$ such that $dt dE=\left[\frac{dt_0}{1+z}\right] \left[(1+z) dE_0\right]$ where $t_0$ and $E_0$ are the time and energy respectively at $z =0$) is given by,
\begin{align}
dN_{\gamma}=& e^{-\tau} \left[(1+z)^{3}\int dM \frac{dn}{dM}(M,z)\times \frac{d\mathcal{N}_{\gamma}}{dE}(E,M,z) \right] \nonumber\\
& \frac{dVdA}{4 \pi (R_{0}S_{k}(r))^2} dE_{0}dt_{0}.
\label{eq_flxeg}
\end{align}
In the above, the volume element $dV$ is given by
\begin{equation}
dV = \frac {\left(R_0 S_k(r)\right)^2 R_0} {(1+z)^3} dr d\Omega_{\rm{detector}},
\end{equation} 
where $S_k(r)$ is Universe's spatial curvature appearing in Robertson-Walker metric. 
The quantity $\frac{dn}{dM}(M,z)$ 
is the halo mass function where as $\frac{d\mathcal{N}_{\gamma}}{dE}(E,M,z)$ is 
the photon energy flux. In this case, we consider that the $\gamma$-rays are originated as the end product of the dark matter annihilation. Therefore computation of dark matter cross-section is important for the calculation of $\frac{d\mathcal{N}_{\gamma}}{dE}$. 
The extragalactic $\gamma$-rays produced at a redshift $z$ suffers 
attenuation during its passage
through intergalactic medium. This attenuation of extragalactic $\gamma$-ray is due to 
the absorption of high energy $\gamma$-rays by extragalactic background light (EBL). Detailed 
studies for this attenuation are given in \citet{cirelli} and Fig.~\ref{fig:tau} is generated 
following \citet{cirelli}. This attenuation can be described 
by an exponential function in terms of the optical depth $\tau$
as $e^{- \tau (z, E_0)}$, $E_0$ being the energy at detection at $z=0$. 
The optical depth is related to the pair production of baryonic 
matter, photon-photon scattering in ambient photon background radiation
(PBR) and photon-photon pair production \cite{cirelli}. In Fig.~\ref{fig:tau} we have shown the dependence of the attenuation factor on redshift ($z$) and the
energy $E_0$ at detection ($z=0$).

The PBR depends on the Cosmic Microwave Background (CMB), the intergalactic stellar light 
and the secondary Infrared (IR) radiation.
The Ultraviolet (UV) background can be originated from intergalactic stellar light. These stellar light may come from the massive and hot stars that were ignited at very low redshift. The two models of UV background are generally used for the background estimation. One is the ``no UV" case where the contribution of the UV is absent, while the other is ``relativistic UV". The latter has been considered in blazar study and it prescribed a certain value for the UV background proton density. But this value is lower than the values estimated in many of the other earlier analyses 
\cite{Franceschini:2008tp}. In this work, a significant amount of contribution of the 
UV background has been taken into account as described in 
\cite{Dominguez:2010bv,Franceschini:2008tp}.
\begin{figure}
	\centering
	\includegraphics[width=0.5\textwidth] {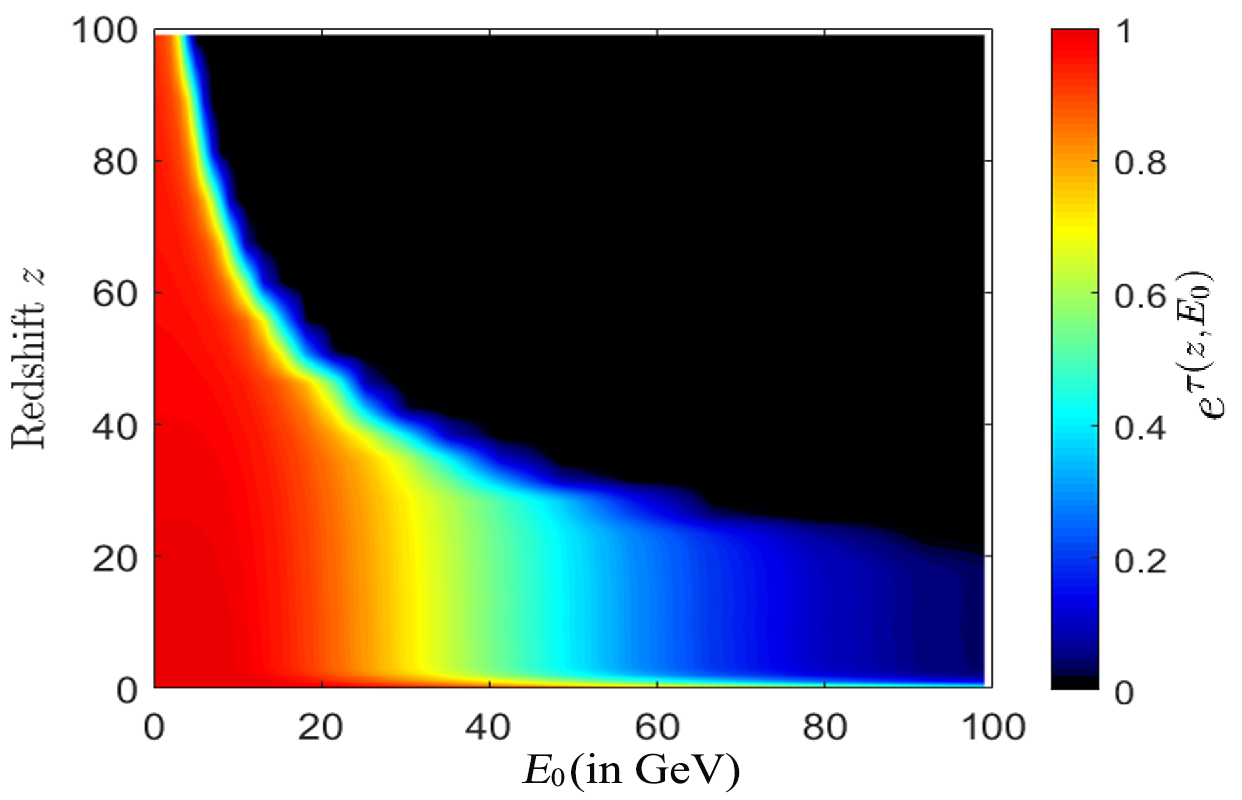}
	\caption{Variation of optical depth $e^{\tau}$ is described as 
		function of energy $E_0$ and redshift $z$. The numerical values of 
		the $e^{\tau}$ is described in the colourbar. \label{fig:tau}}
\end{figure} The diffuse 
extragalactic $\gamma$-ray flux due to DM annihilation is written as,
\begin{align}
\frac{d\phi_{\gamma}}{dE_0}=&\frac{dN_{\gamma}}{dA d\Omega dt_0 dE_0}\nonumber \\
=&\frac{c}{4 \pi} \int dz \frac{e^{-\tau(z,E_0)}}{H_0 h(z)}
\int dM \frac{dn}{dM}(M,z) \nonumber \\
&\frac{d{\cal N}_{\gamma}}{dE}\left(E_0(1+z),M,z\right),
\label{eq:ex_flx_1}
\end{align}
where $c$ is the speed of light in vacuum, $H_0$ denotes the Hubble constant 
at the present epoch and $M$ is the dark matter halo mass. For 
spatially flat Universe ($\Omega_k = 0$), $h(z)=\sqrt{\Omega_m(1+z)^3+\Omega_\Lambda}$, where $\Omega_i\,(i=m,\Lambda, k)$ represents the density parameter for matter ($M$) or dark energy ($\Lambda$) or curvature ($k$). The halo mass function $\frac{dn}{dM}$ is expressed in terms of the fluctuation, the overdensity in structure formation etc. Denoting $\sigma^2(M)$ to be the variance of the linear density field (rms density $=\sigma$) the mass function $f(\sigma)$ extrapolated to redshift $z$ can be written following Press-Schechter model \cite{ex_dndm} as,
\begin{equation}
f(\sigma) = \sqrt{\frac{2}{\pi}} \frac{\delta_{c}}{\sigma} \exp{\left(\frac{-\delta_{c}^2}{2 \sigma^2}\right)}.
\label{neq}
\end{equation}
This expression arises out of the following assumption. After smoothening the linear density perturbations over a mass scale $M$, if in a fraction of space this smooth density field exceeds a threshold $\delta_{c}$ then this fraction of space collapses with mass greater than $M$. This $\delta_{c}$ is called critical overdensity for collapse \footnote{$f$ is also defined as $f(\sigma,z)=\frac{M}{\rho_0}\frac{dn(M,z)}{d \ln \sigma^{-1}}$ where $n$ signifies the halo abundance with mass $<M$ at redshift $z$ and $\rho_0$ be the mean density of the Universe at that redshift \cite{jenkin2001}}. In other word, this is the critical value of initial overdensity that is required for collapse at $z$. This mass function $f$ is also written in terms of a quantity $\nu$ where $\nu$ $(=\delta_{c}/\sigma)$ is the overdensity in units of rms density $\sigma$. The ratio $\nu$ is related to mean square mass fluctuation $\sigma^2(M)$ that is also caused by the non-linear growth of fluctuation. The distribution $f(\nu)$ is the distribution of mass in isolated halos at a given epoch and is related to number densities of halos \footnote{Press and Schechter proposed an ellipsoidal collapse model where the aspects of non spherical collapse are also addressed along with the spherical collapse. In this scenario, the critical overdensity ($\delta_{\rm{sc}}$) for spherical collapse is replaced by ($\delta_{\rm{ec}}$) the same for ellipsoidal collapse. These are related as $\delta_{\rm{ec}}(\sigma,z)=\delta_{\rm{sc}}\left(z\left(1+\beta\left(\frac{\sigma^2}{\delta_{\rm{sc}}^2(z)}\right)^{\gamma}\right)\right)$ with $\beta=0.47$ and $\gamma=0.615$ \cite{Sheth_2001}. For massive objects however $\sigma/\delta_{\rm{sc}}<1$ and $\delta_{\rm{ec}}(\sigma,z)\simeq \delta_{\rm{sc}}(\sigma,z)$}.
The mass density function $\frac{dn}{dM}(M,z)$ is written as \cite{ex_dndm},
\begin{equation}
\frac{dn}{dM} = \frac{{\rho}_{0,m}}{M^2}
\nu f(\nu) \frac{d \log\nu}{d \log M},
\label{eq:massfunc} 
\end{equation}
where ${\rho}_{0, m}$ is the matter density of the comoving background (${\rho}_{0, m}=\rho_c \Omega_m(1+z)^3$, $\rho_c$ is the critical density of the Universe), the ratio $\nu = \delta_{c}/\sigma(M)$ as discussed, where $\delta_{c}$ ($\simeq 1.686$, \cite{jenkin2001,wjp2001}) is the critical overdensity for spherical collapse and $\sigma^2(M)$ is the variance of density fluctuations of a sphere containing mass $M$ ($M \simeq (4/3)\pi R^3 \rho_c (z_c)$ for collapse halos, $R$ being the comoving length and $z_c$ is the redshift at which the halo collapses). The term $\sigma^2 (M)$ can be represented in terms of the power spectrum $P(k)$ of the initial density perturbation as \cite{Sheth},
\begin{equation}
\sigma^2(M) = \frac{1}{2\pi^2}\int_0^{\infty} d^3k \tilde{W}^2(kR) P(k).
\end{equation}
\begin{figure*}
	\centering
	\begin{tabular}{cc}
		\includegraphics[trim=45 265 75 286, clip,width=0.45\textwidth]{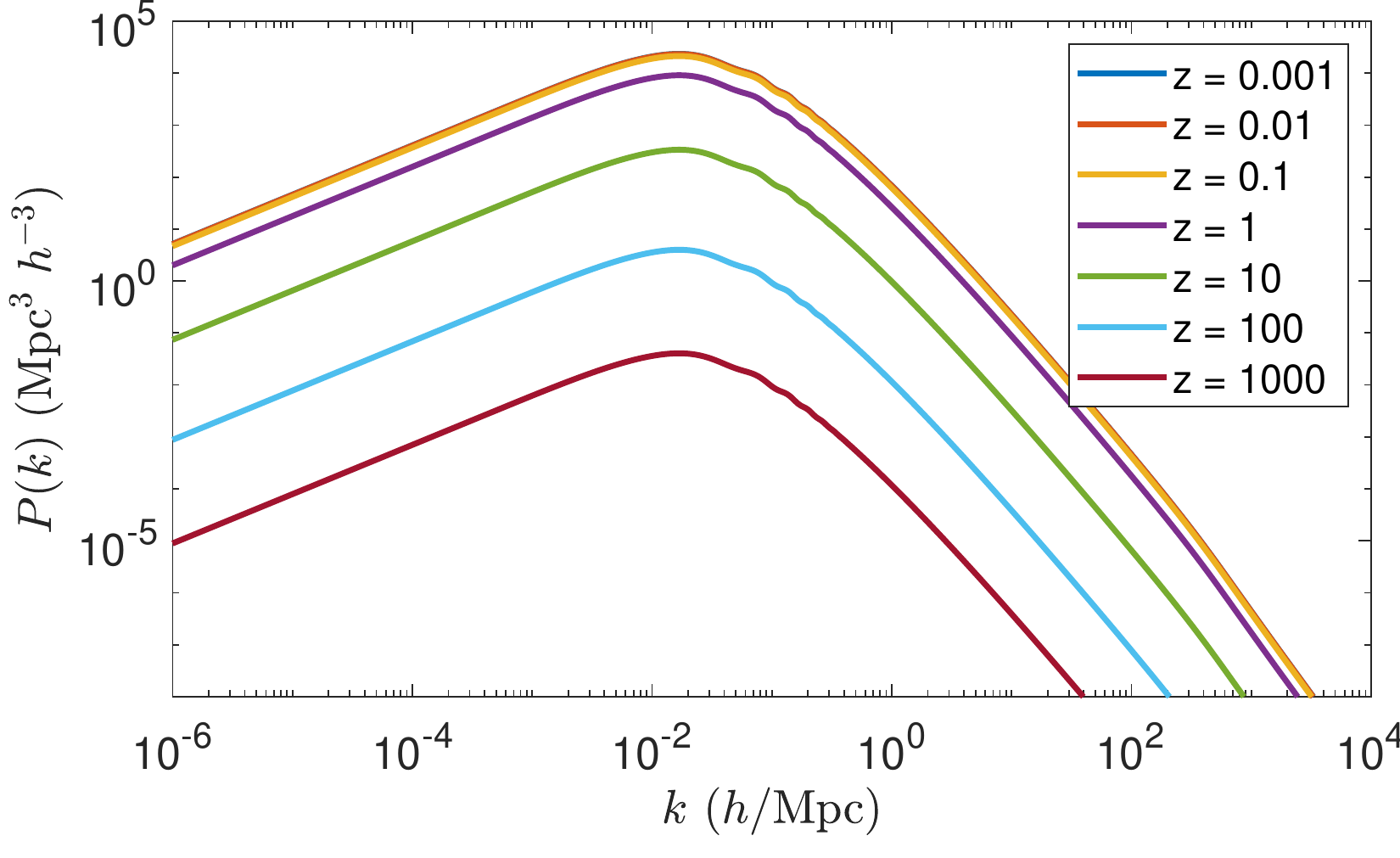} & \includegraphics[trim=70 265 70 286, clip,width=0.45\textwidth]{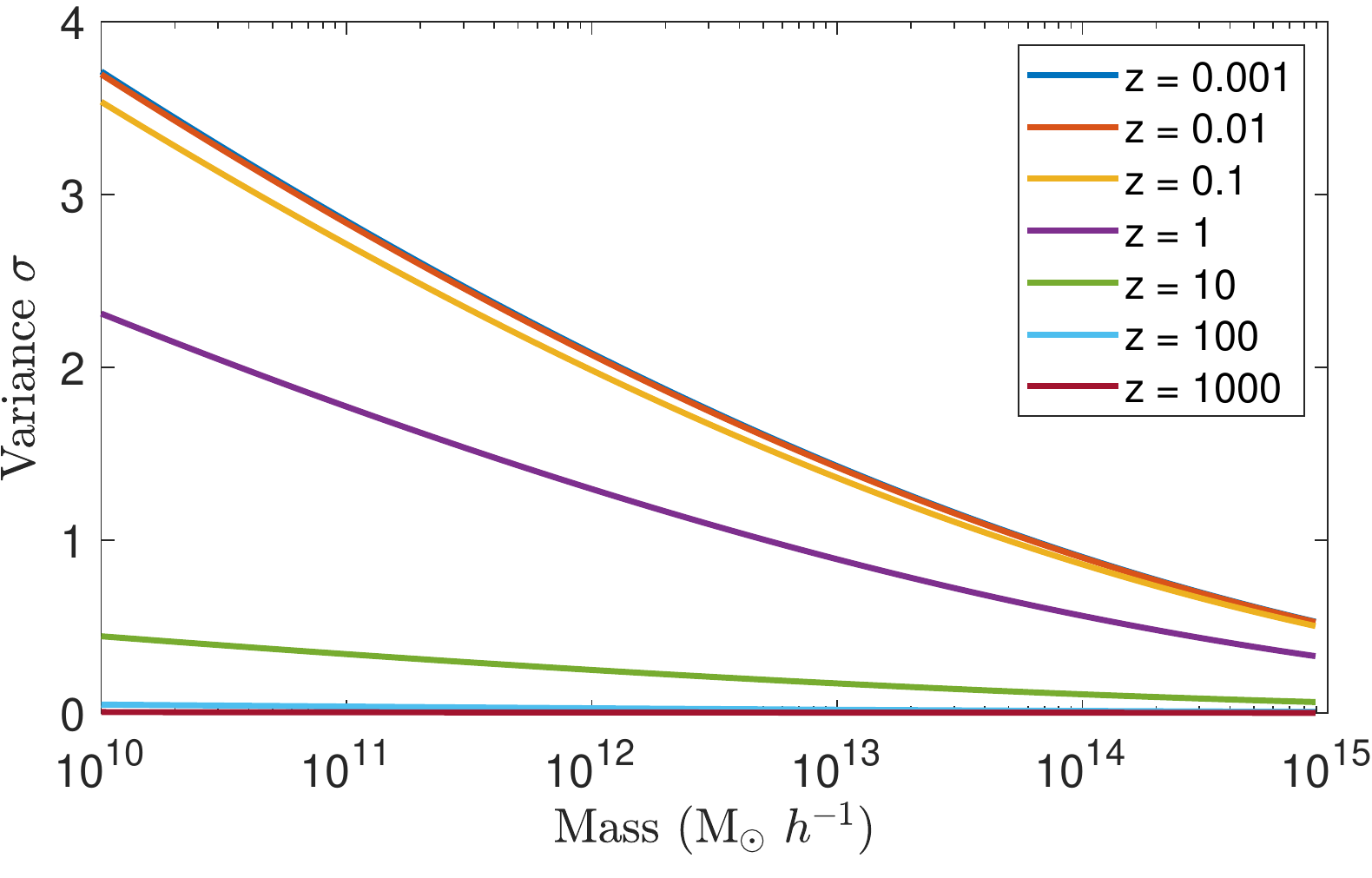} \\
		(a) & (b)\\ 
		\includegraphics[trim=41 265 55 286, clip,width=0.45\textwidth]{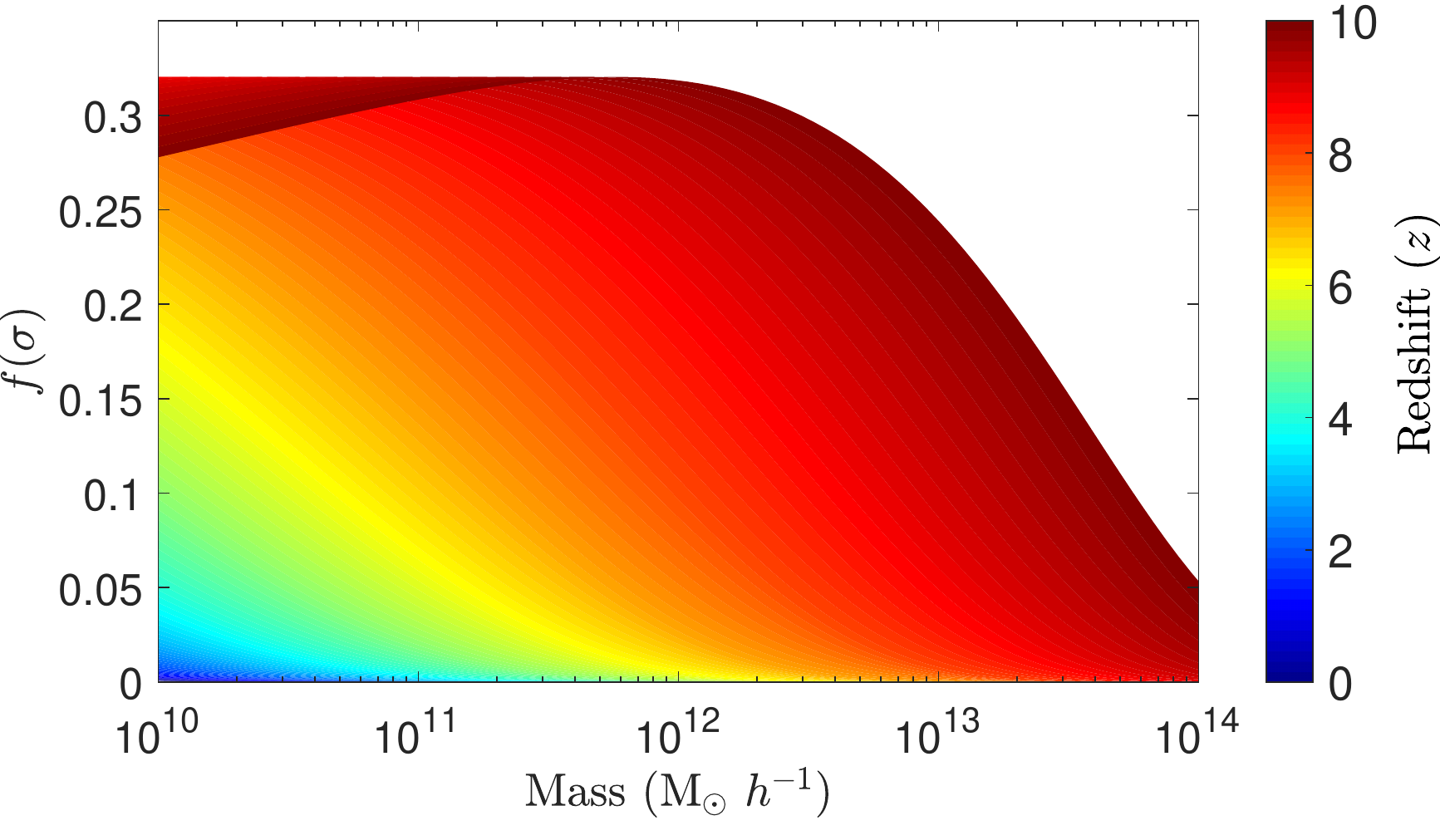} & \includegraphics[trim=35 265 55 286, clip,width=0.45\textwidth]{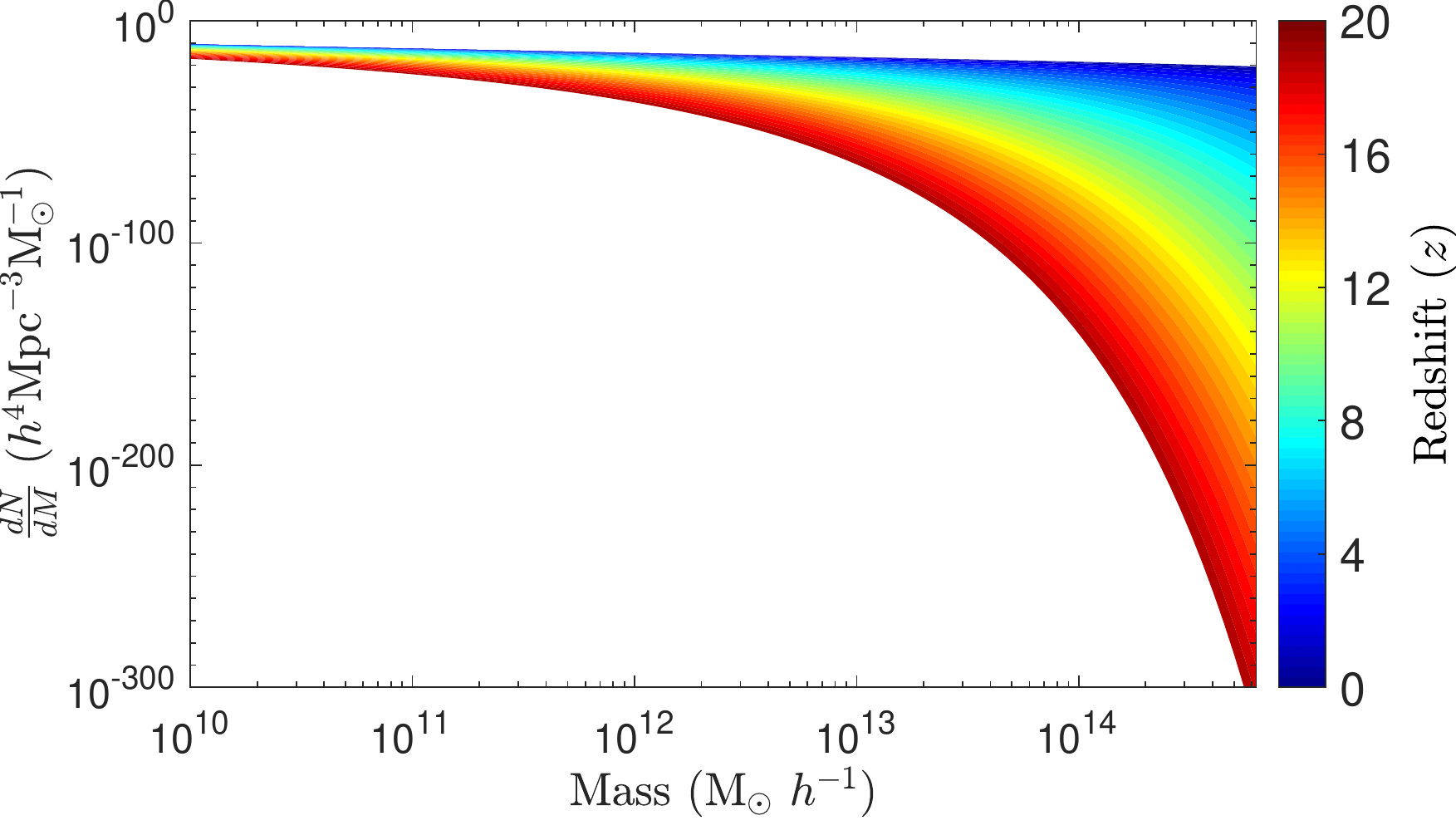}\\
		(c) & (d)\\
	\end{tabular}
	\caption{(a) Fraction of mass collapsed ($f(\sigma)$) for different 
		redshifts $z$ and halo masses $M$ according to the Sheth–Torman model. 
		(b) Variation of $\frac{dN}{dM}$ with halo mass $M$ for different 
		redshift $z$. (c) Variance $\sigma$ of the density perturbations with 
		halo mass for different redshifts ($z$). (d) Variation of the linear power 
		spectrum $P(k)$ of matter density perturbations with the wave 
		number $k$ for different redshifts ($z$).	\label{fig:hmfcalc}}
\end{figure*}
In the above, $\tilde{W}(kR)$ is the Fourier 
transform of the real space top hat window function of radius $R$ \footnote{$\tilde{W}(kR)=3 \dfrac{(\frac{\sin{kR}}{kR}-\cos{kR})}{(kR)^2}$}. The power spectrum is parameterized as $P(k) \propto k^n T^2(k)$ where $n$ is the spectral index and $T$ is a transfer function related to the DM and baryon density in the Universe. Cosmic microwave background data will be useful for its computation. In Fig.~\ref{fig:hmfcalc}a we show the variation of $P(k)$ with wave number $k$ for different $z$ values. We also compute how the variance $\sigma$ varies with halo mass $M$. These variations are plotted in Fig.~\ref{fig:hmfcalc}b for the same set of $z$ values as in Fig.~\ref{fig:hmfcalc}a. The multiplicity function $f(\nu)$ in Eq.~\ref{eq:massfunc} is computed using the relation \cite{Sheth},
\begin{equation}
	\nu f(\nu) = 2 A \left(1+\frac{1}{\nu'^{2p}}\right)
	\left(\frac{\nu'^{2}}{2\pi}\right)^{1/2}
	\exp\left(-\frac{\nu'^{2}}{2}\right);
	\label{eq:nufnu}
\end{equation}
where $\nu'=\sqrt{a}\nu$. Fitting the Eq.~\ref{eq:massfunc} with 
$N$-body simulation of Virgo consortium~\cite{Jenkins:1997en} the 
numerical values of $a$($= 0.707$) and $p = 0.3$ can be obtained. The value of the parameter $A$ in the above equation is adopted as $A=0.322$ \cite{Sheth}. In terms of $\sigma$ the mass function $f(\sigma)$ is written as (with $\nu=\delta_c/\sigma(M)$) 
\begin{equation}
	f(\sigma) = A \sqrt{\displaystyle\frac {2 a} 
		{\pi}} \left [ 1 + \left ( \displaystyle\frac {\sigma^2} 
	{a \delta_c^2} \right)^{P} \right ] \displaystyle\frac {\delta_c} 
	{a} \exp \left [ - \displaystyle\frac {a \delta_c^2} 
	{2 \sigma^2} \right ].
	\label{neqq}
\end{equation}

The function $f(\nu)$ as well as the numerical values for $\nu$ can be 
computed by using Eq.~\ref{eq:nufnu}. The variations of the mass collapse 
function $(f(\sigma))$ in the ellipsoidal models with the halo mass 
$M$ for several values of redshift $z$ (0-10) are demonstrated in Fig.~\ref{fig:hmfcalc}c. Fig.~\ref{fig:hmfcalc}d describes the variations of the considered 
halo mass function $\displaystyle\frac {dn} {dM}$ of 
Sheth-Torman model \cite{Sheth} with 
redshift $z$ and the halo mass $M$. Note that, Fig.~\ref{fig:tau} is just a 
demonstrative plot for the variation of optical depth with energy and 
redshift. In order to generate this demonstrative plot, all the necessary numerical 
calculations have been executed by performing 
HMFcalc \cite{Murray:2013qza} code. 

According to the $\Lambda$CDM cosmological model, the DM halos are formed 
in the bottom-up sequence. In this approach, initially the small clumps 
of matter forms in the presence of a tiny density fluctuation zones 
having a very high gravitational impact. This small scale 
structures grow into the larger ones, eventually forming the larger 
scale structures like the DM halos. The DM density profile of a DM halo 
as per the suggestion of the N-body simulation can be written 
as $\rho(r) = \rho(s) g(r/r_s)$, where $r_s$ and $\rho_s$ indicate 
the scale radius and the scale density for a particular halo model 
respectively. For the halo profile we have chosen NFW halo profile 
\cite{Navarro:1995iw, Navarro:1996gj} depending on which we 
can explain the nature of the function $g(r/r_s)$ (NFW halo profile has 
been mentioned in Table~\ref{app:haloprofile}). The mass of any DM halo contained within 
the radius $r_h$ is given as 
\begin{equation}
	M_h = 4 \pi \rho_s r_h^3 f(r_s/r_h),
	\label{mass}
\end{equation}
where $f(x) = x^3 [ln(1+x)^{-1} - (1+x)^{-1}]$. The NFW profile has two parameters namely a characteristic inner radius $r_s$ and a characteristic inner density $\rho_s$ \cite{Wechsler_2002}. One of these characteristic parameter can be replaced by virial radius or virial mass where, virial mass $M_{\rm{vir}}$ is 
\begin{equation}
	M_h = M_{\rm{vir}} = \displaystyle\frac {4\pi} {3} \Delta_{\rm vir} \bar{\rho} (z) 
	R_{\rm vir}^3,
	\label{mass1}
\end{equation}
where $\Delta_{\rm vir} \bar{\rho} (z)$ is the mean density in the virial radius $R_{\rm{vir}}$ and $\Delta_{\rm vir}$ is the critical over density at virialisation. In the above, $\bar{\rho} (z)$ is the mean Universal density.
For the flat Universe ($\Omega_k = 0$), $\Delta_{\rm{vir}}(z)$ takes the form \cite{Bryan_1998}
\begin{equation}
\Delta_{\rm{vir}} \simeq (18\pi^2 + 82d-39 d^2),
\end{equation}
with $d\equiv d(z) = \frac{\Omega_m(1+z)^3}{(\Omega_m(1+z)^3 + 
\Omega_{\Lambda})} -1$
$\Bigg(d(z) \equiv \Omega(z) - 1 = \displaystyle\frac {\Omega_m (1+z)^3} {E(z)^2} -1,\; \textrm{where}\; E(z) = \displaystyle\frac {H(z)} {H_0}\Bigg)$.

The $\gamma$-ray energy spectrum $\frac {d {\cal N}} {dE}$ depends on 
the halo profile which is taken to be NFW profile in the present calculations.
The shape of the profile can be alternatively described in terms of 
a concentration parameter. As the name suggests this parameter is 
about the concentration of matter in the halo at different positions
and hence is an effective alternative for the description of 
the shape of the halo density profile. In general, the concentration
parameter is formally expressed in terms of the virial radius $R_{\rm vir}$
as $c_{\rm vir} = \frac {R_{\rm vir}} {r^{(-2)}_s}$ where $r^{(-2)}_s$
is the radius at which the logarithmic slope of the density profile
is $-2$ $\left(\frac{d \log (\rho)} {dr} = -2\right)$ \cite{klypin}. Considering the characteristic radius 
$r_s$ of the halo to be the 
radius $r^{(-2)}_s$ and defining 
$x = \frac {r} {r^{(-2)}_s}$, the NFW density profile (Table~\ref{app:haloprofile})
takes the form $\rho (r) = \rho_s g(r/r_s) = \frac {\rho_s}
{x(1+x^2)}$. In the present computation, an $r$-dependent form
	is adopted as $c_{\rm vir} r_{-2} = \frac {R_{\rm vir}} {r}$,
	where $r_{-2} = \frac {r^{(-2)}_s} {r_s}$. With this the $\gamma$-ray energy spectrum 
$\frac{d{\cal N}_{\gamma}}{dE}\left(E_0\,(1+z),M,z\right)$ 
for the $\gamma$-ray (Eq.~\ref{eq:ex_flx_1}) (induced by the dark matter annihilation with annhilation cross-section $\langle \sigma v \rangle$) emitted from a halo of 
mass $M$ at redshift $z$ can be written as,
\begin{align}
\frac{d{\cal N}_{\gamma}}{dE} (E,M,z)=&\frac{\langle\sigma v\rangle}{2}
\frac{dN_{\gamma}(E)}{dE} \int dc^{\prime}_{\rm{vir}}{\cal{P}}
(c^{\prime}_{\rm{vir}})\left(\frac{\rho^{\prime}}{M_{\chi}}\right)^2 \nonumber \\
&\int d^3r g^2(r/a).
\label{eq:dnde_ex}
\end{align}
Here, the differential $\gamma$-ray spectrum is $\frac{dN_{\gamma}(E)}{dE}$ 
and $c_{\rm{vir}}$ is known as the concentration parameter whose lognormal
distribution around a mean value (within 1$\sigma$ \cite{Sheth}) for halos with mass $M$ is denoted as ${\cal{P}}(c_{\rm{vir}})$. We finally have, 
\begin{align}
\frac{d{\cal N}_{\gamma}}{dE} (E,M,z) =& \frac{\sigma v}{2} 
\frac{dN_{\gamma}(E)}{dE}\frac{M}{M_{\chi}^2} 
\frac{\Delta_{\rm{vir}}\bar{\rho} (z)}{3}\nonumber \\
&\int dc^{\prime}_{\rm{vir}} 
{\cal{P}}(c^{\prime}_{\rm{vir}})\frac{(c^{\prime}_{\rm{vir}}r_{-2})^3}
{\left[I_1(c^{\prime}_{\rm{vir}}\,r_{-2})\right]^2} I_2(x_{min},
c^{\prime}_{\rm{vir}}r_{-2}).
\label{eq:dnde2_ex}
\end{align} 
The integration $I_n(x_{min},x_{max})$ is given by 
$I_n(x_{min},x_{max}) = \int_{x_{min}}^{x_{max}} dx\, x^2 g^n(x)$.
Finally the extragalactic $\gamma$-ray flux from DM annihilation takes the 
form~\cite{Ullio:2002pj}
\begin{align}
\frac{d\phi_{\gamma}}{dE_0} =& \frac{\sigma v}{8 \pi} \frac{c}{H_0}
\frac{{\rho}_0^2}{M_{\chi}^2} \int dz (1+z)^3 \frac{\Delta^2(z)}{h(z)}
\frac{dN_{\gamma}(E_0 (1+z))}{dE}\nonumber \\
& e^{-\tau(z,E_0)},
\label{eq:flux2_ex}
\end{align}
with 
\begin{equation}
\Delta^2(z) \equiv \int dM \frac{\nu(z,M) f\left(\nu(z,M)\right)}{\sigma(M)}
\left|\frac{d\sigma}{dM}\right| \Delta_M^2(z,M)\; 
\label{eq:D2}
\end{equation}
and $\left(c_{\rm{vir}} = \frac{R_{\rm{vir}}}{r_s^{(-2)}}\right)$,
\begin{align}
\Delta_M^2(z,M)\equiv&
\frac{\Delta_{\rm{vir}}(z)}{3}\,\int dc^{\,\prime}_{\rm{vir}}\; 
{\cal{P}}(c^{\,\prime}_{\rm{vir}}) \nonumber \\
&\frac{I_2(x_{min},c^{\,\prime}_{\rm{vir}}(z,M)\,r_{-2})}
{\left[I_1(x_{min},c^{\,\prime}_{\rm{vir}}(z,M)\,r_{-2})\right]^2}(c^{\,\prime}_{\rm{vir}}(z,M)\,r_{-2})^3.
\label{eq:D2M}
\end{align}

Two forms for concentration parameter $c_{\rm{vir}}$ are adopted 
for the present computation of extragalactic $\gamma$-ray flux. 
The first form is 
$c_{\rm{vir}}(M,z)=k_{200} 
\left(\mathcal{H}(z_f(M))/\mathcal{H}(z)\right)^{2/3}$ from 
Macci\`{o} et al.\cite{Maccio} with 
$k_{200} \simeq 3.9$, $\mathcal{H}(z)=H(z)/H_0$ and $z_c(M)$ is the
effective redshift when a halo with mass $M$ is formed.
The second form 
$c_{\rm{vir}}(M,z)=6.5\, \mathcal{H}(z)^{-2/3}$
$(M/M_*)^{-0.1}$, $M_*=3.37\times 10^{12} h^{-1}M_\odot$ follows
from a power law model (~\cite{Neto:2007vq, Maccio}).
In what follows we refer this second form for $C_{\rm vir}$ as 
``Power law model for $C_{\rm vir}$" while the former form for 
$C_{\rm vir}$ as ``Macci\`{o} et al. model for $C_{\rm vir}$". 
The dark matter substructure within a halo may form bound subhalos. 
The minimum mass for such subhalos are denoted by $M_{\rm min}$. 
This minimum mass $M_{\rm min}$ for such subhalos are determined 
from the decoupling temperature of dark matter. 
Two values of minimum subhalo mass 
namely $M_{\rm min} =10^{-6} M_{\odot}$ and
$10^{-9}\, M_{\odot}$ \cite{Martinez:2009jh, Bringmann:2009vf} are 
chosen. 

We use Eqs.~\ref{eq_flxeg}-\ref{eq:D2} to compute the extragalactic $\gamma$-ray flux 
(Eq. 18) induced by annihilation of dark matter. As mentioned, in order to 
explore the possibilities that the extragalactic $\gamma$-rays 
could be indirect dark matter signal, the calculations are performed
for each of the two particle dark matter candidates followed from the two particle
dark matter models considered in this work (similar to what has been described in 
Section~\ref{sec:d_gal} for the case pf dwarf galaxies). We mention here that for 
the dark matter models, we adopt exactly the same framework and the numerical values 
of the parameters (such as couplings) as used for the computation related to the 
$\gamma$-ray fluxes of the dwarf galaxies described in Section~\ref{sec:flux}.
While one is a 50 GeV dark matter $-$ the WIMP component
of a two component WIMP-FImP dark matter model (Model I) the other (Model II)
is a 900 GeV Kaluza-Klein (KK) dark matter inspired by  
extra dimensional models (see Section~\ref{sec:intro}).

\begin{figure*}
	\centering
	\includegraphics[width=\textwidth] {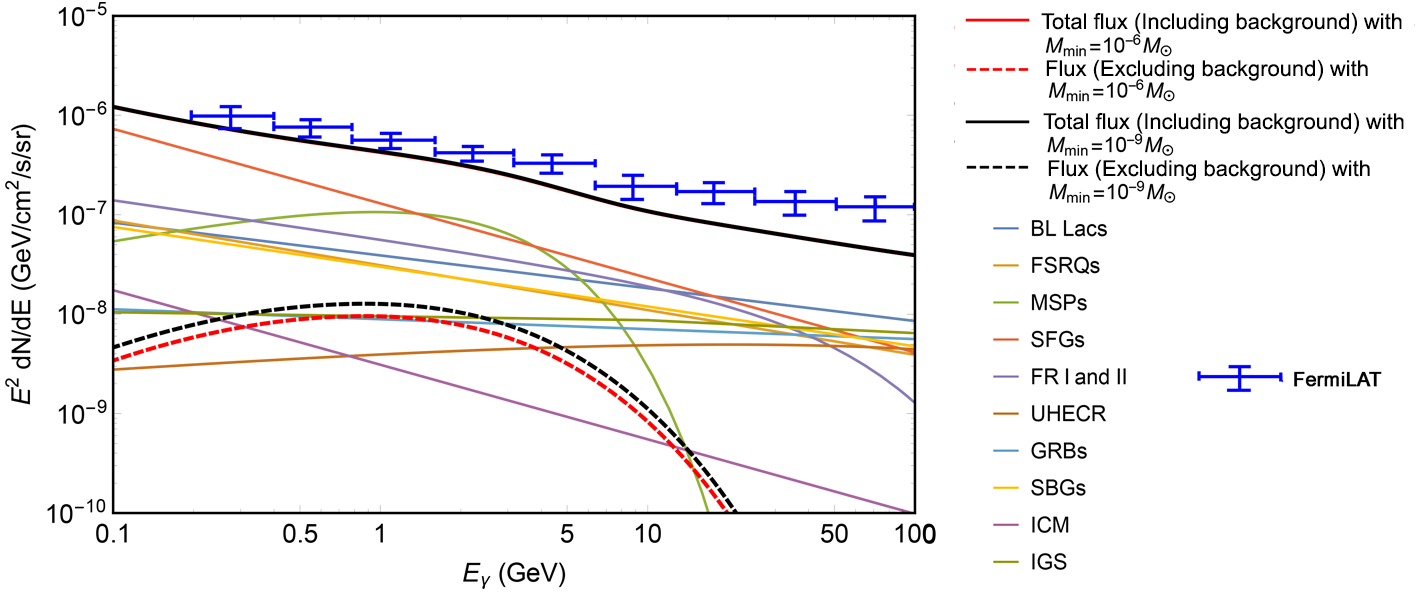}
	\caption{Observed extragalactic $\gamma$-ray fluxes by {\it Fermi}-LAT compared with the total $\gamma$-ray fluxes obtained from the DM annihilation for Model-I DM and other possible non-DM $\gamma$-rays extragalactic sources. For the flux calculation, we have taken into account concentration parameter ($c_{\rm{vir}}$), which is adopt from Macci\`{o} et al. See text for details.\label{fig:maccio_ex}}
\end{figure*}

\begin{figure*}
	\centering
	\includegraphics[width=\textwidth] {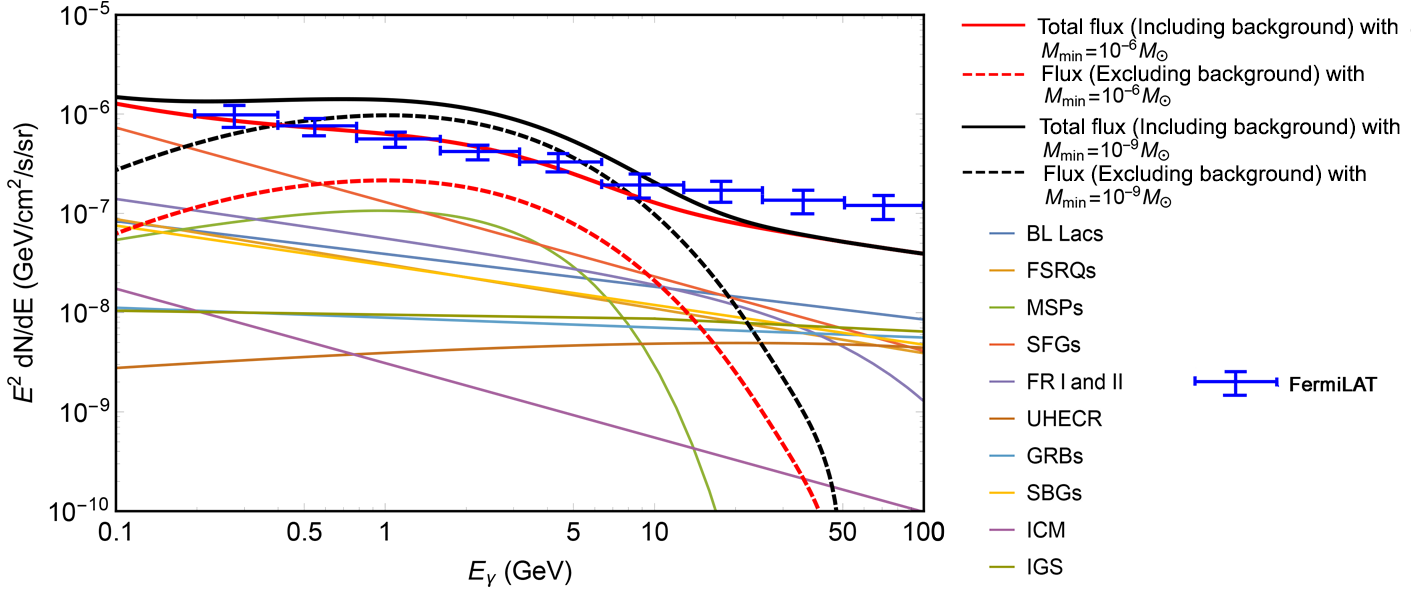}
	\caption{Observed extragalactic $\gamma$-ray fluxes by {\it Fermi}-LAT compared with the total $\gamma$-ray fluxes obtained from the DM annihilation for Model-I DM and other possible non-DM $\gamma$-rays extragalactic sources. In this case the power law for $c_{\rm{vir}}$ is used for the computation of flux. See text for details.\label{fig:power_ex}}
\end{figure*}

\begin{figure*}
	\centering
	\includegraphics[width=\textwidth] {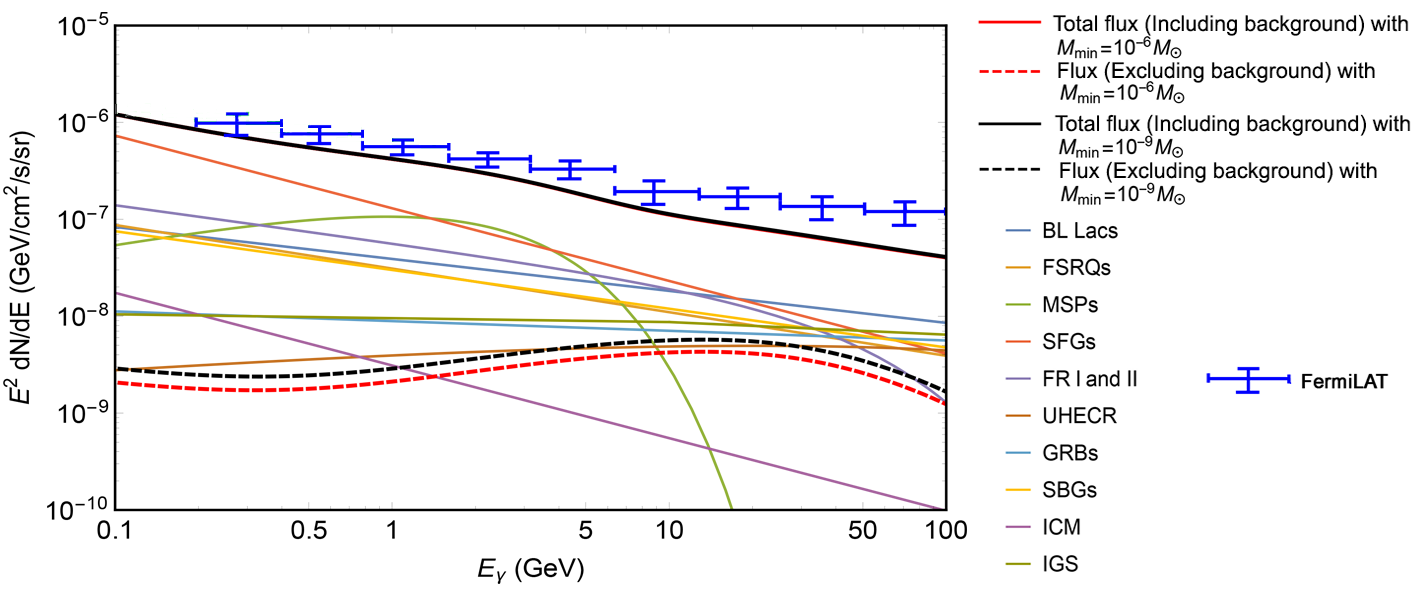}
	\caption{Same as Fig.~\ref{fig:maccio_ex} but for the DM candidate of Model II. See text for details.\label{fig:maccio_ex_kk}}
\end{figure*}

\begin{figure*}
	\centering
	\includegraphics[width=\textwidth] {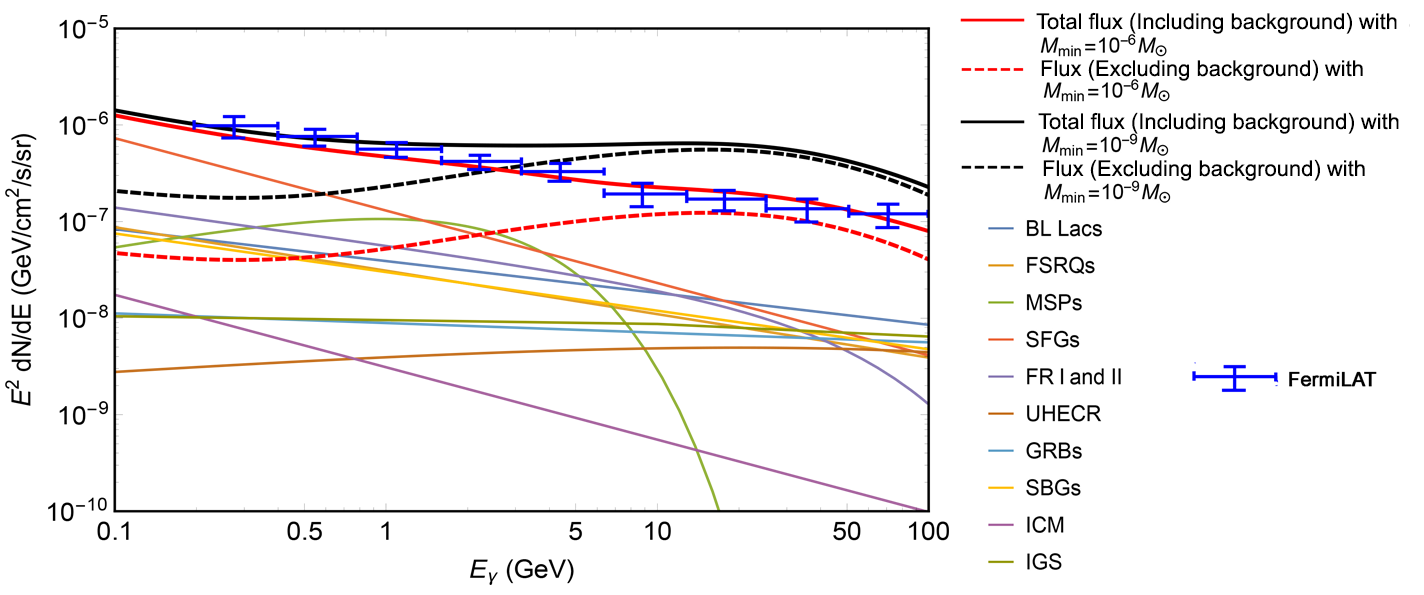}
	\caption{Same as Fig.~\ref{fig:power_ex} but for the DM candidate of Model II. See text for details.\label{fig:power_ex_kk}}
\end{figure*}

We also estimate the background flux from different possible extragalactic 
astrophysical sources. The diffuse $\gamma$-ray background may include contributions from
BL Lacertea objects (BL Lacs), flat spectrum radio quasars (FSRQs),
millisecond pulsars (MSPs), star forming galaxy (SFG),
Fanarof-Riley (FR) radio galaxies of type I (FRI) and type II (FRII),
ultra high energy cosmic rays (UHECRs), $\gamma$-ray bursts (GRBs),
star burst galaxy (SBG), ultra high energy protons interacting with the inter-cluster
material (UHEp ICM) and
gravitationally induced shock waves (IGS). These along with the
nature of their empirical nature
(power spectra \cite{Tavakoli:2013zva}) are tabulated 
in Table~\ref{tab:power_sp}.

The sum total of the calculated $\gamma$-ray flux and the background in 
case of two dark matter models considered are shown in Figs.~\ref{fig:maccio_ex}-\ref{fig:power_ex_kk}.
While the computed results with Model I are given in Figs.~\ref{fig:maccio_ex}-\ref{fig:power_ex}, 
in Figs.~\ref{fig:maccio_ex_kk}-\ref{fig:power_ex_kk} we plot the results for Model II. In all the figures, 
however the extragalactic backgrounds from each of the possible 
non-DM sources (Table~\ref{tab:power_sp}) are shown.     
The results computed with each of the two chosen values of minimum 
subhalo mass ($M_{\rm min} = 10^{-6} M_\odot$ and 
$M_{\rm min} = 10^{-9} M_\odot$) are shown in Fig.~\ref{fig:maccio_ex} and 
Fig.~\ref{fig:power_ex} for the case of Model I and in Fig.~\ref{fig:maccio_ex_kk} and Fig.\ref{fig:power_ex_kk}  
when computations are made for the KK dark matter (Model II).

In all the Figs.~\ref{fig:maccio_ex}-\ref{fig:power_ex_kk}, the sum total of the computed extragalactic 
$\gamma$-ray flux 
(assumed to have originated from dark matter annihilation) and the 
background contributions (Table~\ref{tab:power_sp}) are plotted for DM Model I and 
DM Model II as be the case. The observational results of {\it Fermi}-LAT \cite{2010PhRvL.104j1101A,Ackermann:2014usa}
experiment is also shown for comparison. It is seen from Figs.~\ref{fig:maccio_ex}-\ref{fig:power_ex_kk} that for both Model I and Model II of dark matter, the sum 
total of calculated flux (both with $M_{\rm min} = 10^{-6}M_\odot$ 
and  $M_{\rm min} = 10^{-9}M_\odot$) and background 
is always lower than the {\it Fermi}-LAT results for both the 
chosen values of $M$ when the 
``Macci\`{o} et al. model for $C_{\rm vir}$" is used for the calculation 
of $\gamma$-ray flux from DM annihilation (Fig.~\ref{fig:maccio_ex} and Fig.~\ref{fig:maccio_ex_kk}) and hence 
no $\gamma$-ray signal 
from possible extragalactic DM annihilation can be detected above the 
background. The fluxes only due to the dark matter annihilation 
(no background) computed with each of the two chosen values of $M_{\rm min}$ 
are also shown in Figs.~\ref{fig:maccio_ex}-\ref{fig:power_ex_kk} by dashed plots. 
For all the cases considered in Figs.~\ref{fig:maccio_ex}-\ref{fig:power_ex_kk} for DM candidates in 
Model I and Model II, it is observed that the calculated 
flux with $M_{\rm min} = 10^{-9}M_\odot$ always lie above than
when computed with $M_{\rm min} = 10^{-6}M_\odot$.
From Fig.~\ref{fig:power_ex} and Fig.~\ref{fig:power_ex_kk}, it is seen that the  
computed $\gamma$-ray flux (added with the background from non-DM 
sources) with ``Power law model for $C_{\rm vir}$" and 
$M_{\rm min} = 10^{-9}M_\odot$ (solid black line) goes beyond the 
{\it Fermi}-LAT data upto around $E_\gamma = 10$ GeV for Model I (Fig.~\ref{fig:power_ex}).
Even only the computed flux (without the background) with 
$M_{\rm min} = 10^{-9}M_\odot$ goes beyond the {\it Fermi}-LAT data within certain $\gamma$-energy range (Fig.~\ref{fig:power_ex}).   

Similar trends are also seen for the case of KK dark matter also
(Fig.~\ref{fig:maccio_ex_kk} and Fig.~\ref{fig:power_ex_kk}). From Fig.~\ref{fig:maccio_ex_kk}, one sees that when 
``Macci\`{o} et al. model for $C_{\rm vir}$" is considered the total 
computed flux (including non-DM background) always lies below 
the observed flux by {\it Fermi}-LAT in the considered 
range for $E_\gamma$. But here also the flux with 
$M_{\rm min} = 10^{-9}M_\odot$
are closer to the observed results than  when computed with 
$M_{\rm min} = 10^{-6}M_\odot$. Interesting results are obtained 
for KK dark matter when ``Power law 
model for $C_{\rm vir}$" is used in the calculations (Fig.~\ref{fig:power_ex_kk}). In Fig.~\ref{fig:power_ex_kk}
one observes that the total calculated flux (including the background)
with $M_{\rm min} = 10^{-6}M_\odot$ agrees very well with the {\it Fermi}-LAT
results for almost the whole considered range of $E_\gamma$ upto the energy $\sim 1$ GeV. 
The total flux, when $M_{\rm min} = 10^{-9}M_\odot$, agrees with {\it Fermi}-LAT in lower 
energy region (upto $\sim1$ GeV). Comparing with Fig.~\ref{fig:power_ex}
(similar case for Model I)
it appears that the extragalactic $\gamma$-rays from the annihilation of
KK dark matter in extra dimensional model better agrees with experimental
results than the WIMP DM of Model I with dark matter mass of 50 GeV.
It can also be observed from Figs.~\ref{fig:maccio_ex}-\ref{fig:power_ex_kk} that 
``Power law model for $C_{\rm vir}$", the concentration parameter,
is more suited than the  
``Macci\`{o} et al. model for $C_{\rm vir}$" in the present calculations. 

\begin{table*}
	\centering
	\caption{The contributions of non-dark matter sources to the extragalactic $\gamma$ ray background. \label{tab:power_sp}}
	\begin{tabular}{l c}
		\hline
		Non-DM source & $\frac{dN}{dE}$ in $\rm{GeV^{-1}cm^{-2}s^{-1}sr^{-1}}$\\ \hline
		BL Lacs &$3.9\times10^{-8}E_{\gamma}^{-2.23}$\\
		FSRQ &$3.1\times10^{-8}E_{\gamma}^{-2.45}$\\
		MSP &$1.8\times10^{-7}E_{\gamma}^{-1.5} \exp{\left(-\frac{E_{\gamma}}{1.9}\right)}$\\
		SFG &$1.3\times10^{-7}E_{\gamma}^{-2.75}$\\
		FR I and FR II &$5.7\times10^{-8}E_{\gamma}^{-2.39} \exp{\left(-\frac{E_{\gamma}}{50.0}\right)}$\\
		UHECR & $4.8\times10^{-9}E_{\gamma}^{-1.8}\exp{\left[-\left(\frac{E_{\gamma}}{100.0}\right)^{0.35}\right]}$\\
		GRB &$8.9\times10^{-9}E_{\gamma}^{-2.1}$\\
		SBG &$0.3\times10^{-7}E_{\gamma}^{-2.4}$\\
		UHEp ICM &$3.1\times10^{-9}E_{\gamma}^{-2.75}$\\
		IGS &$0.87\times10^{-10} \times \begin{array}{ll}
		\left(\frac{E_{\gamma}}{10.0}\right)^{-2.04}&~~~\rm{for} ~~E_{\gamma}<10\rm{GeV}\\
		\left(\frac{E_{\gamma}}{10.0}\right)^{-2.13}&~~~\rm{for} ~~E_{\gamma}>10\rm{GeV}\\
		\end{array}$\\
		\hline
	\end{tabular}
\end{table*}

\section{Summary and Discussions}\label{summ}

In this work, we explore the observational upper limits of $\gamma$-ray flux
from 45 dwarf spheroidal galaxies and relate these to the $\gamma$-rays
that could be produced from annihilation of dark matter in dSphs. From the
mass to luminosity ratios, it appears that dSphs could be rich in dark matter and the
dark matter can produce $\gamma$-rays via the annihilation process. For our
analysis, we consider two dark matter candidates in two particle dark matter models. One is a two
component WIMP-FImP model of which the WIMP component undergoes
annihilation to produce the $\gamma$-ray flux. The WIMP component is a Dirac
singlet fermion and its additional U(1)$_{\rm DM}$ charge prevents its
interaction with Standard Model (SM) fermions. But the interaction between the WIMP fermion and the SM sector can be occurred via the higgs portal. The benchmark mass for this dark matter is chosen
to be 50 GeV for the present analysis. The other particle dark matter
chosen for the analysis is Kaluza-Klein (KK)
dark matter inspired by models of extra dimensions.
In Universal extra dimensional model $B^1$, which we have consider as the KK dark matter candidate in this work, is the KK partner of hypercharge guage boson. This is stable due to the conservation of KK parity $(-1)^{KK}$ where KK is the KK number of the Kaluza-Klein tower related to the extra dimensional momentum.
The mass of $B^1$ that satisfies the dark matter relic density is $\sim 900$ GeV. In this work, we have taken the mass of $B^1$ to be 900 GeV which is much
higher than the Higgs portal fermionic dark matter in Model I.
It appears from the analysis that for both the Higgs portal model and
KK model, the dark matter annihilations to $\gamma$-rays for 45 dwarf galaxies
are well within the observational upper bounds of the $\gamma$-ray flux
for all the 45 galaxies considered. While the Higgs portal dark matter
(Model I) covers a shorter range, the Kaluza-Klein dark matter
having higher mass range can probe the $\gamma$-ray flux at higher energy
range.

We have also extended our analyses for the case of possible extragalactic
signature of $\gamma$-rays from dark matter annihilations.
If detected, such signals could be the indirect extragalactic
dark matter signals. For this extragalactic case also, we
adopt the same dark matter models and the model parameter values 
given in Tables~\ref{tab:model1} and \ref{tab:model} and discussed 
in Sections~\ref{sec:intro} and \ref{sec:flux}.
But there can be many extragalactic  $\gamma$-ray sources 
other than possible dark matter annihilations. These non-DM 
sources for extragalactic $\gamma$-rays originate the background for 
the $\gamma$-rays produced via possible extragalactic dark matter 
annihilations. We have made an estimation of the flux for 
such extragalactic sources. The $\gamma$-ray flux from dark matter 
annihilations primarily depends on the annihilation cross-section 
of the particle dark matter candidate. In addition to this, the parameters which we need to taken into account for the $\gamma$-ray flux calculation are the dark matter halo mass function, 
the density fluctuation in the halo, the 
linear and non-linear growth of density perturbation and their collapse, 
the virial radius, the minimum mass 
$M_{\rm min}$ required for the formation of the subhalo within a DM halo,
the $\gamma$-ray spectrum 
($\frac {dN} {dE}$), the attenuation factor of these $\gamma$-rays  
during it's passage towards a terrestrial detector etc. 
It is also required to use a feasible model for dark matter 
halo density profile. In this calculations, NFW density profile 
has been considered and this NFW density profile is a function of concentration parameter ($c_{\rm vir}$), which plays a major role to compute the extragalactic $\gamma$-ray flux originated from the dark matter annihilation. For the analysis, we adopt two forms for $c_{\rm vir}$, where
one is a power form and the other one is a form given by
\cite{Maccio}. We have considered two distinct values foe minimum mass $M_{\rm min}$, which are
$M_{\rm min} = 10^{-6} M_\odot$ and  
$M_{\rm min} = 10^{-9} M_\odot$. 
The calculations are performed for the case of 
the fermionic WIMP dark matter in Model I and the Kaluza Klein extra 
dimensional dark matter candidate in Model II. 
The computed results are then compared
with the observed results for {\it Fermi}-LAT satellite 
borne experiments. Our analyses show that 
the power law choice for $c_{\rm vir}$ yields better results in comparison to the other choice. For the case of power law choice for $c_{\rm vir}$,
the $\gamma$-ray flux from the annihilation of fermionic 
WIMP dark matter of mass 50 GeV in Model I compares well with {\it Fermi}-LAT
data at least upto $E_\gamma \sim 10$ GeV when $M_{\rm min} = 10^{-6} M_\odot$
and lies above the {\it Fermi}-LAT (upto $E_\gamma \sim 10$ GeV) 
when $M_{\rm min} = 10^{-9} M_\odot$. Better agreements are obtained
for the Kaluza Klein dark matter candidate of Universal Extra Dimension
model (Model II in this work). The KK dark matter candidate is more massive ($\sim$900 GeV) than the WIMP dark matter candidate in Model I ($\sim$50 GeV).
The computed flux for this KK DM candidate (for power law choice
of $c_{\rm vir}$) agrees satisfactorily with the 
{\it Fermi}-LAT results for $M_{\rm min} = 10^{-6} M_\odot$ having a wider energy 
range than in case of the former DM candidate.

The present analyses therefore demonstrate the 
possibilities of detecting indirect signal of dark matter from extragalactic
origins as well as from dwarf spheroidals. The results also indicate
that the particle nature of dark matter can be probed from the 
study of the $\gamma$-rays from both dwarf galaxies and extragalactic 
sources.     

\section*{Acknowledgements}

	Two of the authors (S.B. and A.H.) wish to acknowledge the support received from St. Xavier’s College, Kolkata. One of the authors (A.H.) also acknowledges the University Grant Commission (UGC) of the Government of India, for providing financial support, in the form of UGC-CSIR NET-JRF. One of the authors (MP) thanks the DST-INSPIRE fellowship  (DST/INSPIRE/FELLOWSHIP/IF160004) grant by DST, Govt. of India.

\section*{Data Availability}

	The data underlying this article are available in the article.


\bibliographystyle{aipauth4-1}
\bibliography{pub_mnras} 

\begin{thebibliography}{60}%
\makeatletter
\providecommand \@ifxundefined [1]{%
 \@ifx{#1\undefined}
}%
\providecommand \@ifnum [1]{%
 \ifnum #1\expandafter \@firstoftwo
 \else \expandafter \@secondoftwo
 \fi
}%
\providecommand \@ifx [1]{%
 \ifx #1\expandafter \@firstoftwo
 \else \expandafter \@secondoftwo
 \fi
}%
\providecommand \natexlab [1]{#1}%
\providecommand \enquote  [1]{``#1''}%
\providecommand \bibnamefont  [1]{#1}%
\providecommand \bibfnamefont [1]{#1}%
\providecommand \citenamefont [1]{#1}%
\providecommand \href@noop [0]{\@secondoftwo}%
\providecommand \href [0]{\begingroup \@sanitize@url \@href}%
\providecommand \@href[1]{\@@startlink{#1}\@@href}%
\providecommand \@@href[1]{\endgroup#1\@@endlink}%
\providecommand \@sanitize@url [0]{\catcode `\\12\catcode `\$12\catcode
  `\&12\catcode `\#12\catcode `\^12\catcode `\_12\catcode `\%12\relax}%
\providecommand \@@startlink[1]{}%
\providecommand \@@endlink[0]{}%
\providecommand \url  [0]{\begingroup\@sanitize@url \@url }%
\providecommand \@url [1]{\endgroup\@href {#1}{\urlprefix }}%
\providecommand \urlprefix  [0]{URL }%
\providecommand \Eprint [0]{\href }%
\providecommand \doibase [0]{http://dx.doi.org/}%
\providecommand \selectlanguage [0]{\@gobble}%
\providecommand \bibinfo  [0]{\@secondoftwo}%
\providecommand \bibfield  [0]{\@secondoftwo}%
\providecommand \translation [1]{[#1]}%
\providecommand \BibitemOpen [0]{}%
\providecommand \bibitemStop [0]{}%
\providecommand \bibitemNoStop [0]{.\EOS\space}%
\providecommand \EOS [0]{\spacefactor3000\relax}%
\providecommand \BibitemShut  [1]{\csname bibitem#1\endcsname}%
\let\auto@bib@innerbib\@empty
\bibitem [{\citenamefont {{Abdo}}\ \emph {et~al.}(2010)\citenamefont {{Abdo}}
  \emph {et~al.}}]{2010PhRvL.104j1101A}%
  \BibitemOpen
  \bibfield  {author} {\bibinfo {author} {\bibnamefont {{Abdo}}, \bibfnamefont
  {A.~A.}} \emph {et~al.},\ }\href {\doibase 10.1103/PhysRevLett.104.101101}
  {\bibfield  {journal} {\bibinfo  {journal} {\prl}\ }\textbf {\bibinfo
  {volume} {104}},\ \bibinfo {eid} {101101} (\bibinfo {year} {2010})},\ \Eprint
  {http://arxiv.org/abs/1002.3603} {arXiv:1002.3603 [astro-ph.HE]} \BibitemShut
  {NoStop}%
\bibitem [{\citenamefont {Ackermann}\ \emph {et~al.}(2015)\citenamefont
  {Ackermann} \emph {et~al.}}]{Ackermann:2015tah}%
  \BibitemOpen
  \bibfield  {author} {\bibinfo {author} {\bibnamefont {Ackermann},
  \bibfnamefont {M.}} \emph {et~al.} (\bibinfo {collaboration} {Fermi-LAT}),\
  }\href {\doibase 10.1088/1475-7516/2015/09/008} {\bibfield  {journal}
  {\bibinfo  {journal} {JCAP}\ }\textbf {\bibinfo {volume} {1509}},\ \bibinfo
  {pages} {008} (\bibinfo {year} {2015})},\ \Eprint
  {http://arxiv.org/abs/1501.05464} {arXiv:1501.05464 [astro-ph.CO]}
  \BibitemShut {NoStop}%
\bibitem [{\citenamefont {{Ackermann}}\ \emph {et~al.}(2015)\citenamefont
  {{Ackermann}} \emph {et~al.}}]{fermilat}%
  \BibitemOpen
  \bibfield  {author} {\bibinfo {author} {\bibnamefont {{Ackermann}},
  \bibfnamefont {M.}} \emph {et~al.},\ }\href {\doibase
  10.1103/PhysRevLett.115.231301} {\bibfield  {journal} {\bibinfo  {journal}
  {Physical Review Letters}\ }\textbf {\bibinfo {volume} {115}},\ \bibinfo
  {eid} {231301} (\bibinfo {year} {2015})},\ \Eprint
  {http://arxiv.org/abs/1503.02641} {arXiv:1503.02641 [astro-ph.HE]}
  \BibitemShut {NoStop}%
\bibitem [{\citenamefont {Ackermann}\ \emph {et~al.}(2015)\citenamefont
  {Ackermann} \emph {et~al.}}]{Ackermann:2014usa}%
  \BibitemOpen
  \bibfield  {author} {\bibinfo {author} {\bibnamefont {Ackermann},
  \bibfnamefont {M.}} \emph {et~al.} (\bibinfo {collaboration} {Fermi-LAT}),\
  }\href {\doibase 10.1088/0004-637X/799/1/86} {\bibfield  {journal} {\bibinfo
  {journal} {Astrophys. J.}\ }\textbf {\bibinfo {volume} {799}},\ \bibinfo
  {pages} {86} (\bibinfo {year} {2015})},\ \Eprint
  {http://arxiv.org/abs/1410.3696} {arXiv:1410.3696 [astro-ph.HE]} \BibitemShut
  {NoStop}%
\bibitem [{\citenamefont {{Ajello}}\ \emph {et~al.}(2015)\citenamefont
  {{Ajello}} \emph {et~al.}}]{Ajello:2015mfa}%
  \BibitemOpen
  \bibfield  {author} {\bibinfo {author} {\bibnamefont {{Ajello}},
  \bibfnamefont {M.}} \emph {et~al.},\ }\href {\doibase
  10.1088/2041-8205/800/2/L27} {\bibfield  {journal} {\bibinfo  {journal} {The
  Astrophysical Journal Letters}\ }\textbf {\bibinfo {volume} {800}},\ \bibinfo
  {eid} {L27} (\bibinfo {year} {2015})},\ \Eprint
  {http://arxiv.org/abs/1501.05301} {arXiv:1501.05301 [astro-ph.HE]}
  \BibitemShut {NoStop}%
\bibitem [{\citenamefont {Albert}\ \emph {et~al.}(2017)\citenamefont {Albert}
  \emph {et~al.}}]{Fermi-LAT:2016uux}%
  \BibitemOpen
  \bibfield  {author} {\bibinfo {author} {\bibnamefont {Albert}, \bibfnamefont
  {A.}} \emph {et~al.} (\bibinfo {collaboration} {Fermi-LAT, DES}),\ }\href
  {\doibase 10.3847/1538-4357/834/2/110} {\bibfield  {journal} {\bibinfo
  {journal} {Astrophys. J.}\ }\textbf {\bibinfo {volume} {834}},\ \bibinfo
  {pages} {110} (\bibinfo {year} {2017})},\ \Eprint
  {http://arxiv.org/abs/1611.03184} {arXiv:1611.03184 [astro-ph.HE]}
  \BibitemShut {NoStop}%
\bibitem [{\citenamefont {Ando}(2005)}]{Ando_2005}%
  \BibitemOpen
  \bibfield  {author} {\bibinfo {author} {\bibnamefont {Ando}, \bibfnamefont
  {S.}},\ }\href {\doibase 10.1103/PhysRevLett.94.171303} {\bibfield  {journal}
  {\bibinfo  {journal} {Phys. Rev. Lett.}\ }\textbf {\bibinfo {volume} {94}},\
  \bibinfo {pages} {171303} (\bibinfo {year} {2005})}\BibitemShut {NoStop}%
\bibitem [{\citenamefont {Appelquist}, \citenamefont {Cheng},\ and\
  \citenamefont {Dobrescu}(2001)}]{ued2}%
  \BibitemOpen
  \bibfield  {author} {\bibinfo {author} {\bibnamefont {Appelquist},
  \bibfnamefont {T.}}, \bibinfo {author} {\bibnamefont {Cheng}, \bibfnamefont
  {H.-C.}}, \ and\ \bibinfo {author} {\bibnamefont {Dobrescu}, \bibfnamefont
  {B.~A.}},\ }\href {\doibase 10.1103/PhysRevD.64.035002} {\bibfield  {journal}
  {\bibinfo  {journal} {Phys. Rev.}\ }\textbf {\bibinfo {volume} {D64}},\
  \bibinfo {pages} {035002} (\bibinfo {year} {2001})},\ \Eprint
  {http://arxiv.org/abs/hep-ph/0012100} {arXiv:hep-ph/0012100 [hep-ph]}
  \BibitemShut {NoStop}%
\bibitem [{\citenamefont {{Atwood}}\ \emph {et~al.}(2009)\citenamefont
  {{Atwood}} \emph {et~al.}}]{fermilatold}%
  \BibitemOpen
  \bibfield  {author} {\bibinfo {author} {\bibnamefont {{Atwood}},
  \bibfnamefont {W.~B.}} \emph {et~al.},\ }\href {\doibase
  10.1088/0004-637X/697/2/1071} {\bibfield  {journal} {\bibinfo  {journal} {The
  Astrophysical Journal}\ }\textbf {\bibinfo {volume} {697}},\ \bibinfo {pages}
  {1071} (\bibinfo {year} {2009})},\ \Eprint {http://arxiv.org/abs/0902.1089}
  {arXiv:0902.1089 [astro-ph.IM]} \BibitemShut {NoStop}%
\bibitem [{\citenamefont {Banik}\ and\ \citenamefont
  {Majumdar}(2015)}]{BANIK2015420}%
  \BibitemOpen
  \bibfield  {author} {\bibinfo {author} {\bibnamefont {Banik}, \bibfnamefont
  {A.~D.}}\ and\ \bibinfo {author} {\bibnamefont {Majumdar}, \bibfnamefont
  {D.}},\ }\href {\doibase https://doi.org/10.1016/j.physletb.2015.03.003}
  {\bibfield  {journal} {\bibinfo  {journal} {Physics Letters B}\ }\textbf
  {\bibinfo {volume} {743}},\ \bibinfo {pages} {420 } (\bibinfo {year}
  {2015})}\BibitemShut {NoStop}%
\bibitem [{\citenamefont {Bergstr{\"o}m}, \citenamefont {Edsj{\"o}},\ and\
  \citenamefont {Ullio}(2001)}]{Bergstrom:2001jj}%
  \BibitemOpen
  \bibfield  {author} {\bibinfo {author} {\bibnamefont {Bergstr{\"o}m},
  \bibfnamefont {L.}}, \bibinfo {author} {\bibnamefont {Edsj{\"o}},
  \bibfnamefont {J.}}, \ and\ \bibinfo {author} {\bibnamefont {Ullio},
  \bibfnamefont {P.}},\ }\href@noop {} {\bibfield  {journal} {\bibinfo
  {journal} {Physical Review Letters}\ }\textbf {\bibinfo {volume} {87}},\
  \bibinfo {pages} {251301} (\bibinfo {year} {2001})}\BibitemShut {NoStop}%
\bibitem [{\citenamefont {Biswas}, \citenamefont {Majumdar},\ and\
  \citenamefont {Roy}(2015)}]{Biswas:2015sva}%
  \BibitemOpen
  \bibfield  {author} {\bibinfo {author} {\bibnamefont {Biswas}, \bibfnamefont
  {A.}}, \bibinfo {author} {\bibnamefont {Majumdar}, \bibfnamefont {D.}}, \
  and\ \bibinfo {author} {\bibnamefont {Roy}, \bibfnamefont {P.}},\ }\href
  {\doibase 10.1007/JHEP04(2015)065} {\bibfield  {journal} {\bibinfo  {journal}
  {JHEP}\ }\textbf {\bibinfo {volume} {04}},\ \bibinfo {pages} {065} (\bibinfo
  {year} {2015})},\ \Eprint {http://arxiv.org/abs/1501.02666} {arXiv:1501.02666
  [hep-ph]} \BibitemShut {NoStop}%
\bibitem [{\citenamefont {Boddy}\ \emph {et~al.}(2018)\citenamefont {Boddy},
  \citenamefont {Kumar}, \citenamefont {Marfatia},\ and\ \citenamefont
  {Sandick}}]{jfact1}%
  \BibitemOpen
  \bibfield  {author} {\bibinfo {author} {\bibnamefont {Boddy}, \bibfnamefont
  {K.~K.}}, \bibinfo {author} {\bibnamefont {Kumar}, \bibfnamefont {J.}},
  \bibinfo {author} {\bibnamefont {Marfatia}, \bibfnamefont {D.}}, \ and\
  \bibinfo {author} {\bibnamefont {Sandick}, \bibfnamefont {P.}},\ }\href
  {\doibase 10.1103/PhysRevD.97.095031} {\bibfield  {journal} {\bibinfo
  {journal} {Phys. Rev. D}\ }\textbf {\bibinfo {volume} {97}},\ \bibinfo
  {pages} {095031} (\bibinfo {year} {2018})}\BibitemShut {NoStop}%
\bibitem [{\citenamefont {Bringmann}(2009)}]{Bringmann:2009vf}%
  \BibitemOpen
  \bibfield  {author} {\bibinfo {author} {\bibnamefont {Bringmann},
  \bibfnamefont {T.}},\ }\href {\doibase 10.1088/1367-2630/11/10/105027}
  {\bibfield  {journal} {\bibinfo  {journal} {New Journal of Physics}\ }\textbf
  {\bibinfo {volume} {11}},\ \bibinfo {pages} {105027} (\bibinfo {year}
  {2009})}\BibitemShut {NoStop}%
\bibitem [{\citenamefont {Bringmann}\ \emph {et~al.}(2014)\citenamefont
  {Bringmann}, \citenamefont {Calore}, \citenamefont {Di~Mauro},\ and\
  \citenamefont {Donato}}]{Calore:2013yia}%
  \BibitemOpen
  \bibfield  {author} {\bibinfo {author} {\bibnamefont {Bringmann},
  \bibfnamefont {T.}}, \bibinfo {author} {\bibnamefont {Calore}, \bibfnamefont
  {F.}}, \bibinfo {author} {\bibnamefont {Di~Mauro}, \bibfnamefont {M.}}, \
  and\ \bibinfo {author} {\bibnamefont {Donato}, \bibfnamefont {F.}},\ }\href
  {\doibase 10.1103/PhysRevD.89.023012} {\bibfield  {journal} {\bibinfo
  {journal} {Phys. Rev. D}\ }\textbf {\bibinfo {volume} {89}},\ \bibinfo
  {pages} {023012} (\bibinfo {year} {2014})}\BibitemShut {NoStop}%
\bibitem [{\citenamefont {Bryan}\ and\ \citenamefont
  {Norman}(1998)}]{Bryan_1998}%
  \BibitemOpen
  \bibfield  {author} {\bibinfo {author} {\bibnamefont {Bryan}, \bibfnamefont
  {G.~L.}}\ and\ \bibinfo {author} {\bibnamefont {Norman}, \bibfnamefont
  {M.~L.}},\ }\href {\doibase 10.1086/305262} {\bibfield  {journal} {\bibinfo
  {journal} {The Astrophysical Journal}\ }\textbf {\bibinfo {volume} {495}},\
  \bibinfo {pages} {80} (\bibinfo {year} {1998})}\BibitemShut {NoStop}%
\bibitem [{\citenamefont {Burkert}(1995)}]{burkert1}%
  \BibitemOpen
  \bibfield  {author} {\bibinfo {author} {\bibnamefont {Burkert}, \bibfnamefont
  {A.}},\ }\href {\doibase 10.1086/309560} {\bibfield  {journal} {\bibinfo
  {journal} {The Astrophysical Journal}\ }\textbf {\bibinfo {volume} {447}}
  (\bibinfo {year} {1995}),\ 10.1086/309560}\BibitemShut {NoStop}%
\bibitem [{\citenamefont {Cheng}, \citenamefont {Feng},\ and\ \citenamefont
  {Matchev}(2002)}]{Cheng:2002ej}%
  \BibitemOpen
  \bibfield  {author} {\bibinfo {author} {\bibnamefont {Cheng}, \bibfnamefont
  {H.-C.}}, \bibinfo {author} {\bibnamefont {Feng}, \bibfnamefont {J.~L.}}, \
  and\ \bibinfo {author} {\bibnamefont {Matchev}, \bibfnamefont {K.~T.}},\
  }\href {\doibase 10.1103/PhysRevLett.89.211301} {\bibfield  {journal}
  {\bibinfo  {journal} {Phys. Rev. Lett.}\ }\textbf {\bibinfo {volume} {89}},\
  \bibinfo {pages} {211301} (\bibinfo {year} {2002})},\ \Eprint
  {http://arxiv.org/abs/hep-ph/0207125} {arXiv:hep-ph/0207125 [hep-ph]}
  \BibitemShut {NoStop}%
\bibitem [{\citenamefont {Cholis}, \citenamefont {Hooper},\ and\ \citenamefont
  {McDermott}()}]{Cholis:2013ena}%
  \BibitemOpen
  \bibfield  {author} {\bibinfo {author} {\bibnamefont {Cholis}, \bibfnamefont
  {I.}}, \bibinfo {author} {\bibnamefont {Hooper}, \bibfnamefont {D.}}, \ and\
  \bibinfo {author} {\bibnamefont {McDermott}, \bibfnamefont {S.~D.}},\ }\href
  {\doibase 10.1088/1475-7516/2014/02/014} {\bibfield  {journal} {\bibinfo
  {journal} {Journal of Cosmology and Astroparticle Physics}\
  }10.1088/1475-7516/2014/02/014}\BibitemShut {NoStop}%
\bibitem [{\citenamefont {Cirelli}\ \emph {et~al.}(2011)\citenamefont {Cirelli}
  \emph {et~al.}}]{cirelli}%
  \BibitemOpen
  \bibfield  {author} {\bibinfo {author} {\bibnamefont {Cirelli}, \bibfnamefont
  {M.}} \emph {et~al.},\ }\href {\doibase 10.1088/1475-7516/2011/03/051}
  {\bibfield  {journal} {\bibinfo  {journal} {Journal of Cosmology and
  Astroparticle Physics}\ }\textbf {\bibinfo {volume} {2011}},\ \bibinfo
  {pages} {051} (\bibinfo {year} {2011})}\BibitemShut {NoStop}%
\bibitem [{\citenamefont {Di~Mauro}(2015)}]{DiMauro:2015ika}%
  \BibitemOpen
  \bibfield  {author} {\bibinfo {author} {\bibnamefont {Di~Mauro},
  \bibfnamefont {M.}},\ }in\ \href@noop {} {\emph {\bibinfo {booktitle} {{5th
  International Fermi Symposium Nagoya, Japan, October 20-24, 2014}}}}\
  (\bibinfo {year} {2015})\ \Eprint {http://arxiv.org/abs/1502.02566}
  {arXiv:1502.02566 [astro-ph.HE]} \BibitemShut {NoStop}%
\bibitem [{\citenamefont {Di~Mauro}\ and\ \citenamefont
  {Donato}(2015)}]{DiMauro:2015tfa}%
  \BibitemOpen
  \bibfield  {author} {\bibinfo {author} {\bibnamefont {Di~Mauro},
  \bibfnamefont {M.}}\ and\ \bibinfo {author} {\bibnamefont {Donato},
  \bibfnamefont {F.}},\ }\href {\doibase 10.1103/PhysRevD.91.123001} {\bibfield
   {journal} {\bibinfo  {journal} {Phys. Rev.}\ }\textbf {\bibinfo {volume}
  {D91}},\ \bibinfo {pages} {123001} (\bibinfo {year} {2015})},\ \Eprint
  {http://arxiv.org/abs/1501.05316} {arXiv:1501.05316 [astro-ph.HE]}
  \BibitemShut {NoStop}%
\bibitem [{\citenamefont {{Dom{\'{\i}}nguez}}\ \emph
  {et~al.}(2011)\citenamefont {{Dom{\'{\i}}nguez}} \emph
  {et~al.}}]{Dominguez:2010bv}%
  \BibitemOpen
  \bibfield  {author} {\bibinfo {author} {\bibnamefont {{Dom{\'{\i}}nguez}},
  \bibfnamefont {A.}} \emph {et~al.},\ }\href {\doibase
  10.1111/j.1365-2966.2010.17631.x} {\bibfield  {journal} {\bibinfo  {journal}
  {Mon. Not. R. Astr. Soc.}\ }\textbf {\bibinfo {volume} {410}},\ \bibinfo
  {pages} {2556} (\bibinfo {year} {2011})},\ \Eprint
  {http://arxiv.org/abs/1007.1459} {arXiv:1007.1459} \BibitemShut {NoStop}%
\bibitem [{\citenamefont {Dutta~Banik}\ \emph {et~al.}(2017)\citenamefont
  {Dutta~Banik}, \citenamefont {Pandey}, \citenamefont {Majumdar},\ and\
  \citenamefont {Biswas}}]{pandeymajumdar}%
  \BibitemOpen
  \bibfield  {author} {\bibinfo {author} {\bibnamefont {Dutta~Banik},
  \bibfnamefont {A.}}, \bibinfo {author} {\bibnamefont {Pandey}, \bibfnamefont
  {M.}}, \bibinfo {author} {\bibnamefont {Majumdar}, \bibfnamefont {D.}}, \
  and\ \bibinfo {author} {\bibnamefont {Biswas}, \bibfnamefont {A.}},\ }\href
  {\doibase 10.1140/epjc/s10052-017-5221-y} {\bibfield  {journal} {\bibinfo
  {journal} {The European Physical Journal C}\ }\textbf {\bibinfo {volume}
  {77}},\ \bibinfo {pages} {657} (\bibinfo {year} {2017})}\BibitemShut
  {NoStop}%
\bibitem [{\citenamefont {Einasto}(1965)}]{einasto}%
  \BibitemOpen
  \bibfield  {author} {\bibinfo {author} {\bibnamefont {Einasto}, \bibfnamefont
  {J.}},\ }\href@noop {} {\bibfield  {journal} {\bibinfo  {journal} {Astrofiz.
  Alma-Ata}\ }\textbf {\bibinfo {volume} {51}} (\bibinfo {year}
  {1965})}\BibitemShut {NoStop}%
\bibitem [{\citenamefont {{Franceschini}}, \citenamefont {{Rodighiero}},\ and\
  \citenamefont {{Vaccari}}(2008)}]{Franceschini:2008tp}%
  \BibitemOpen
  \bibfield  {author} {\bibinfo {author} {\bibnamefont {{Franceschini}},
  \bibfnamefont {A.}}, \bibinfo {author} {\bibnamefont {{Rodighiero}},
  \bibfnamefont {G.}}, \ and\ \bibinfo {author} {\bibnamefont {{Vaccari}},
  \bibfnamefont {M.}},\ }\href {\doibase 10.1051/0004-6361:200809691}
  {\bibfield  {journal} {\bibinfo  {journal} {Astronomy and Astrophysics}\
  }\textbf {\bibinfo {volume} {487}},\ \bibinfo {pages} {837} (\bibinfo {year}
  {2008})},\ \Eprint {http://arxiv.org/abs/0805.1841} {arXiv:0805.1841}
  \BibitemShut {NoStop}%
\bibitem [{\citenamefont {{Gao}}, \citenamefont {{Stecker}},\ and\
  \citenamefont {{Cline}}(1991)}]{Gao:1991rz}%
  \BibitemOpen
  \bibfield  {author} {\bibinfo {author} {\bibnamefont {{Gao}}, \bibfnamefont
  {Y.-T.}}, \bibinfo {author} {\bibnamefont {{Stecker}}, \bibfnamefont
  {F.~W.}}, \ and\ \bibinfo {author} {\bibnamefont {{Cline}}, \bibfnamefont
  {D.~B.}},\ }\href@noop {} {\bibfield  {journal} {\bibinfo  {journal}
  {Astronomy and Astrophysics}\ }\textbf {\bibinfo {volume} {249}},\ \bibinfo
  {pages} {1} (\bibinfo {year} {1991})}\BibitemShut {NoStop}%
\bibitem [{\citenamefont {Hooper}\ and\ \citenamefont
  {Goodenough}(2011)}]{Hooper:2010mq}%
  \BibitemOpen
  \bibfield  {author} {\bibinfo {author} {\bibnamefont {Hooper}, \bibfnamefont
  {D.}}\ and\ \bibinfo {author} {\bibnamefont {Goodenough}, \bibfnamefont
  {L.}},\ }\href {\doibase 10.1016/j.physletb.2011.02.029} {\bibfield
  {journal} {\bibinfo  {journal} {Phys. Lett. B}\ }\textbf {\bibinfo {volume}
  {697}},\ \bibinfo {pages} {412} (\bibinfo {year} {2011})},\ \Eprint
  {http://arxiv.org/abs/1010.2752} {arXiv:1010.2752 [hep-ph]} \BibitemShut
  {NoStop}%
\bibitem [{\citenamefont {Hooper}\ and\ \citenamefont
  {Linden}(2011)}]{PhysRevD.84.123005}%
  \BibitemOpen
  \bibfield  {author} {\bibinfo {author} {\bibnamefont {Hooper}, \bibfnamefont
  {D.}}\ and\ \bibinfo {author} {\bibnamefont {Linden}, \bibfnamefont {T.}},\
  }\href {\doibase 10.1103/PhysRevD.84.123005} {\bibfield  {journal} {\bibinfo
  {journal} {Phys. Rev. D}\ }\textbf {\bibinfo {volume} {84}},\ \bibinfo
  {pages} {123005} (\bibinfo {year} {2011})}\BibitemShut {NoStop}%
\bibitem [{\citenamefont {Hooper}\ \emph {et~al.}(2008)\citenamefont {Hooper},
  \citenamefont {Zaharijas}, \citenamefont {Finkbeiner},\ and\ \citenamefont
  {Dobler}}]{Hooper:2007gi}%
  \BibitemOpen
  \bibfield  {author} {\bibinfo {author} {\bibnamefont {Hooper}, \bibfnamefont
  {D.}}, \bibinfo {author} {\bibnamefont {Zaharijas}, \bibfnamefont {G.}},
  \bibinfo {author} {\bibnamefont {Finkbeiner}, \bibfnamefont {D.~P.}}, \ and\
  \bibinfo {author} {\bibnamefont {Dobler}, \bibfnamefont {G.}},\ }\href
  {\doibase 10.1103/PhysRevD.77.043511} {\bibfield  {journal} {\bibinfo
  {journal} {Phys. Rev.}\ }\textbf {\bibinfo {volume} {D77}},\ \bibinfo {pages}
  {043511} (\bibinfo {year} {2008})},\ \Eprint {http://arxiv.org/abs/0709.3114}
  {arXiv:0709.3114 [astro-ph]} \BibitemShut {NoStop}%
\bibitem [{\citenamefont {Jenkins}\ \emph {et~al.}(1998)\citenamefont
  {Jenkins}, \citenamefont {Frenk}, \citenamefont {Pearce}, \citenamefont
  {Thomas}, \citenamefont {Colberg}, \citenamefont {White}, \citenamefont
  {Couchman}, \citenamefont {Peacock}, \citenamefont {Efstathiou},\ and\
  \citenamefont {and}}]{Jenkins:1997en}%
  \BibitemOpen
  \bibfield  {author} {\bibinfo {author} {\bibnamefont {Jenkins}, \bibfnamefont
  {A.}}, \bibinfo {author} {\bibnamefont {Frenk}, \bibfnamefont {C.~S.}},
  \bibinfo {author} {\bibnamefont {Pearce}, \bibfnamefont {F.~R.}}, \bibinfo
  {author} {\bibnamefont {Thomas}, \bibfnamefont {P.~A.}}, \bibinfo {author}
  {\bibnamefont {Colberg}, \bibfnamefont {J.~M.}}, \bibinfo {author}
  {\bibnamefont {White}, \bibfnamefont {S.~D.~M.}}, \bibinfo {author}
  {\bibnamefont {Couchman}, \bibfnamefont {H.~M.~P.}}, \bibinfo {author}
  {\bibnamefont {Peacock}, \bibfnamefont {J.~A.}}, \bibinfo {author}
  {\bibnamefont {Efstathiou}, \bibfnamefont {G.}}, \ and\ \bibinfo {author}
  {\bibnamefont {and}, \bibfnamefont {A.~H.~N.}},\ }\href {\doibase
  10.1086/305615} {\bibfield  {journal} {\bibinfo  {journal} {The Astrophysical
  Journal}\ }\textbf {\bibinfo {volume} {499}},\ \bibinfo {pages} {20}
  (\bibinfo {year} {1998})}\BibitemShut {NoStop}%
\bibitem [{\citenamefont {{Jenkins}}\ \emph {et~al.}(2001)\citenamefont
  {{Jenkins}}, \citenamefont {{Frenk}}, \citenamefont {{White}}, \citenamefont
  {{Colberg}}, \citenamefont {{Cole}}, \citenamefont {{Evrard}}, \citenamefont
  {{Couchman}},\ and\ \citenamefont {{Yoshida}}}]{jenkin2001}%
  \BibitemOpen
  \bibfield  {author} {\bibinfo {author} {\bibnamefont {{Jenkins}},
  \bibfnamefont {A.}}, \bibinfo {author} {\bibnamefont {{Frenk}}, \bibfnamefont
  {C.~S.}}, \bibinfo {author} {\bibnamefont {{White}}, \bibfnamefont
  {S.~D.~M.}}, \bibinfo {author} {\bibnamefont {{Colberg}}, \bibfnamefont
  {J.~M.}}, \bibinfo {author} {\bibnamefont {{Cole}}, \bibfnamefont {S.}},
  \bibinfo {author} {\bibnamefont {{Evrard}}, \bibfnamefont {A.~E.}}, \bibinfo
  {author} {\bibnamefont {{Couchman}}, \bibfnamefont {H.~M.~P.}}, \ and\
  \bibinfo {author} {\bibnamefont {{Yoshida}}, \bibfnamefont {N.}},\ }\href
  {\doibase 10.1046/j.1365-8711.2001.04029.x} {\bibfield  {journal} {\bibinfo
  {journal} {Mon. Not. R. Astr. Soc.}\ }\textbf {\bibinfo {volume} {321}},\
  \bibinfo {pages} {372} (\bibinfo {year} {2001})},\ \Eprint
  {http://arxiv.org/abs/astro-ph/0005260} {arXiv:astro-ph/0005260 [astro-ph]}
  \BibitemShut {NoStop}%
\bibitem [{\citenamefont {Klypin}\ \emph {et~al.}(2016)\citenamefont {Klypin},
  \citenamefont {Yepes}, \citenamefont {Gottlöber}, \citenamefont {Prada},\
  and\ \citenamefont {Heß}}]{klypin}%
  \BibitemOpen
  \bibfield  {author} {\bibinfo {author} {\bibnamefont {Klypin}, \bibfnamefont
  {A.}}, \bibinfo {author} {\bibnamefont {Yepes}, \bibfnamefont {G.}}, \bibinfo
  {author} {\bibnamefont {Gottlöber}, \bibfnamefont {S.}}, \bibinfo {author}
  {\bibnamefont {Prada}, \bibfnamefont {F.}}, \ and\ \bibinfo {author}
  {\bibnamefont {Heß}, \bibfnamefont {S.}},\ }\href {\doibase
  10.1093/mnras/stw248} {\bibfield  {journal} {\bibinfo  {journal} {Mon. Not.
  R. Astr. Soc.}\ }\textbf {\bibinfo {volume} {457}},\ \bibinfo {pages} {4340}
  (\bibinfo {year} {2016})},\ \Eprint
  {http://arxiv.org/abs/https://academic.oup.com/mnras/article-pdf/457/4/4340/18515365/stw248.pdf}
  {https://academic.oup.com/mnras/article-pdf/457/4/4340/18515365/stw248.pdf}
  \BibitemShut {NoStop}%
\bibitem [{\citenamefont {Kong}\ and\ \citenamefont {Matchev}(2007)}]{ued3}%
  \BibitemOpen
  \bibfield  {author} {\bibinfo {author} {\bibnamefont {Kong}, \bibfnamefont
  {K.}}\ and\ \bibinfo {author} {\bibnamefont {Matchev}, \bibfnamefont
  {K.~T.}},\ }\bibfield  {booktitle} {\emph {\bibinfo {booktitle}
  {{Supersymmetry and the unification of fundamental interactions. Proceedings,
  14th International Conference, SUSY 2006, Irvine, USA, June 12-17, 2006}}},\
  }\href {\doibase 10.1063/1.2735221} {\bibfield  {journal} {\bibinfo
  {journal} {AIP Conf. Proc.}\ }\textbf {\bibinfo {volume} {903}},\ \bibinfo
  {pages} {451} (\bibinfo {year} {2007})},\ \Eprint
  {http://arxiv.org/abs/hep-ph/0610057} {arXiv:hep-ph/0610057 [hep-ph]}
  \BibitemShut {NoStop}%
\bibitem [{\citenamefont {Lin}\ and\ \citenamefont {Li}(2019)}]{mnras1}%
  \BibitemOpen
  \bibfield  {author} {\bibinfo {author} {\bibnamefont {Lin}, \bibfnamefont
  {H.-N.}}\ and\ \bibinfo {author} {\bibnamefont {Li}, \bibfnamefont {X.}},\
  }\href {\doibase 10.1093/mnras/stz1698} {\bibfield  {journal} {\bibinfo
  {journal} {Mon. Not. R. Astr. Soc.}\ }\textbf {\bibinfo {volume} {487}},\
  \bibinfo {pages} {5679} (\bibinfo {year} {2019})},\ \Eprint
  {http://arxiv.org/abs/https://academic.oup.com/mnras/article-pdf/487/4/5679/28897927/stz1698.pdf}
  {https://academic.oup.com/mnras/article-pdf/487/4/5679/28897927/stz1698.pdf}
  \BibitemShut {NoStop}%
\bibitem [{\citenamefont {{Macci{\`o}}}, \citenamefont {{Dutton}},\ and\
  \citenamefont {{van den Bosch}}(2008)}]{Maccio}%
  \BibitemOpen
  \bibfield  {author} {\bibinfo {author} {\bibnamefont {{Macci{\`o}}},
  \bibfnamefont {A.~V.}}, \bibinfo {author} {\bibnamefont {{Dutton}},
  \bibfnamefont {A.~A.}}, \ and\ \bibinfo {author} {\bibnamefont {{van den
  Bosch}}, \bibfnamefont {F.~C.}},\ }\href {\doibase
  10.1111/j.1365-2966.2008.14029.x} {\bibfield  {journal} {\bibinfo  {journal}
  {Mon. Not. R. Astr. Soc.}\ }\textbf {\bibinfo {volume} {391}},\ \bibinfo
  {pages} {1940} (\bibinfo {year} {2008})},\ \Eprint
  {http://arxiv.org/abs/0805.1926} {arXiv:0805.1926} \BibitemShut {NoStop}%
\bibitem [{\citenamefont {Majumdar}(2003)}]{Majumdar:2003dj}%
  \BibitemOpen
  \bibfield  {author} {\bibinfo {author} {\bibnamefont {Majumdar},
  \bibfnamefont {D.}},\ }\href {\doibase 10.1142/S0217732303011423} {\bibfield
  {journal} {\bibinfo  {journal} {Mod. Phys. Lett.}\ }\textbf {\bibinfo
  {volume} {A18}},\ \bibinfo {pages} {1705} (\bibinfo {year}
  {2003})}\BibitemShut {NoStop}%
\bibitem [{\citenamefont {Martinez}\ \emph {et~al.}(2009)\citenamefont
  {Martinez}, \citenamefont {Bullock}, \citenamefont {Kaplinghat},
  \citenamefont {Strigari},\ and\ \citenamefont {Trotta}}]{Martinez:2009jh}%
  \BibitemOpen
  \bibfield  {author} {\bibinfo {author} {\bibnamefont {Martinez},
  \bibfnamefont {G.~D.}}, \bibinfo {author} {\bibnamefont {Bullock},
  \bibfnamefont {J.~S.}}, \bibinfo {author} {\bibnamefont {Kaplinghat},
  \bibfnamefont {M.}}, \bibinfo {author} {\bibnamefont {Strigari},
  \bibfnamefont {L.~E.}}, \ and\ \bibinfo {author} {\bibnamefont {Trotta},
  \bibfnamefont {R.}},\ }\href {\doibase 10.1088/1475-7516/2009/06/014}
  {\bibfield  {journal} {\bibinfo  {journal} {JCAP}\ }\textbf {\bibinfo
  {volume} {0906}},\ \bibinfo {pages} {014} (\bibinfo {year} {2009})},\ \Eprint
  {http://arxiv.org/abs/0902.4715} {arXiv:0902.4715 [astro-ph.HE]} \BibitemShut
  {NoStop}%
\bibitem [{\citenamefont {Modak}\ and\ \citenamefont
  {Majumdar}(2015)}]{Modak:2015uda}%
  \BibitemOpen
  \bibfield  {author} {\bibinfo {author} {\bibnamefont {Modak}, \bibfnamefont
  {K.~P.}}\ and\ \bibinfo {author} {\bibnamefont {Majumdar}, \bibfnamefont
  {D.}},\ }\href {\doibase 10.1088/0067-0049/219/2/37} {\bibfield  {journal}
  {\bibinfo  {journal} {Astrophys. J. Suppl.}\ }\textbf {\bibinfo {volume}
  {219}},\ \bibinfo {pages} {37} (\bibinfo {year} {2015})},\ \Eprint
  {http://arxiv.org/abs/1502.05682} {arXiv:1502.05682 [hep-ph]} \BibitemShut
  {NoStop}%
\bibitem [{\citenamefont {Murray}, \citenamefont {Power},\ and\ \citenamefont
  {Robotham}(2013)}]{Murray:2013qza}%
  \BibitemOpen
  \bibfield  {author} {\bibinfo {author} {\bibnamefont {Murray}, \bibfnamefont
  {S.}}, \bibinfo {author} {\bibnamefont {Power}, \bibfnamefont {C.}}, \ and\
  \bibinfo {author} {\bibnamefont {Robotham}, \bibfnamefont {A.}},\ }\href@noop
  {} {\  (\bibinfo {year} {2013})},\ \Eprint {http://arxiv.org/abs/1306.6721}
  {arXiv:1306.6721 [astro-ph.CO]} \BibitemShut {NoStop}%
\bibitem [{\citenamefont {Navarro}, \citenamefont {Frenk},\ and\ \citenamefont
  {White}(1996)}]{Navarro:1995iw}%
  \BibitemOpen
  \bibfield  {author} {\bibinfo {author} {\bibnamefont {Navarro}, \bibfnamefont
  {J.~F.}}, \bibinfo {author} {\bibnamefont {Frenk}, \bibfnamefont {C.~S.}}, \
  and\ \bibinfo {author} {\bibnamefont {White}, \bibfnamefont {S.~D.~M.}},\
  }\href {\doibase 10.1086/177173} {\bibfield  {journal} {\bibinfo  {journal}
  {Astrophys. J.}\ }\textbf {\bibinfo {volume} {462}},\ \bibinfo {pages} {563}
  (\bibinfo {year} {1996})},\ \Eprint {http://arxiv.org/abs/astro-ph/9508025}
  {arXiv:astro-ph/9508025 [astro-ph]} \BibitemShut {NoStop}%
\bibitem [{\citenamefont {{Navarro}}, \citenamefont {{Frenk}},\ and\
  \citenamefont {{White}}(1997)}]{Navarro:1996gj}%
  \BibitemOpen
  \bibfield  {author} {\bibinfo {author} {\bibnamefont {{Navarro}},
  \bibfnamefont {J.~F.}}, \bibinfo {author} {\bibnamefont {{Frenk}},
  \bibfnamefont {C.~S.}}, \ and\ \bibinfo {author} {\bibnamefont {{White}},
  \bibfnamefont {S.~D.~M.}},\ }\href {\doibase 10.1086/304888} {\bibfield
  {journal} {\bibinfo  {journal} {The Astrophysical Journal}\ }\textbf
  {\bibinfo {volume} {490}},\ \bibinfo {pages} {493} (\bibinfo {year}
  {1997})},\ \Eprint {http://arxiv.org/abs/astro-ph/9611107} {astro-ph/9611107}
  \BibitemShut {NoStop}%
\bibitem [{\citenamefont {Neto}\ \emph {et~al.}(2007)\citenamefont {Neto} \emph
  {et~al.}}]{Neto:2007vq}%
  \BibitemOpen
  \bibfield  {author} {\bibinfo {author} {\bibnamefont {Neto}, \bibfnamefont
  {A.~F.}} \emph {et~al.},\ }\href {\doibase 10.1111/j.1365-2966.2007.12381.x}
  {\bibfield  {journal} {\bibinfo  {journal} {Mon. Not. R. Astr. Soc.}\
  }\textbf {\bibinfo {volume} {381}},\ \bibinfo {pages} {1450} (\bibinfo {year}
  {2007})}\BibitemShut {NoStop}%
\bibitem [{\citenamefont {Ng}\ \emph {et~al.}(2014)\citenamefont {Ng},
  \citenamefont {Laha}, \citenamefont {Campbell}, \citenamefont {Horiuchi},
  \citenamefont {Dasgupta}, \citenamefont {Murase},\ and\ \citenamefont
  {Beacom}}]{Ng:2013xha}%
  \BibitemOpen
  \bibfield  {author} {\bibinfo {author} {\bibnamefont {Ng}, \bibfnamefont
  {K.~C.~Y.}}, \bibinfo {author} {\bibnamefont {Laha}, \bibfnamefont {R.}},
  \bibinfo {author} {\bibnamefont {Campbell}, \bibfnamefont {S.}}, \bibinfo
  {author} {\bibnamefont {Horiuchi}, \bibfnamefont {S.}}, \bibinfo {author}
  {\bibnamefont {Dasgupta}, \bibfnamefont {B.}}, \bibinfo {author}
  {\bibnamefont {Murase}, \bibfnamefont {K.}}, \ and\ \bibinfo {author}
  {\bibnamefont {Beacom}, \bibfnamefont {J.~F.}},\ }\href {\doibase
  10.1103/PhysRevD.89.083001} {\bibfield  {journal} {\bibinfo  {journal} {Phys.
  Rev. D}\ }\textbf {\bibinfo {volume} {89}},\ \bibinfo {pages} {083001}
  (\bibinfo {year} {2014})}\BibitemShut {NoStop}%
\bibitem [{\citenamefont {Oda}, \citenamefont {Totani},\ and\ \citenamefont
  {Nagashima}(2005)}]{Oda_2005}%
  \BibitemOpen
  \bibfield  {author} {\bibinfo {author} {\bibnamefont {Oda}, \bibfnamefont
  {T.}}, \bibinfo {author} {\bibnamefont {Totani}, \bibfnamefont {T.}}, \ and\
  \bibinfo {author} {\bibnamefont {Nagashima}, \bibfnamefont {M.}},\ }\href
  {\doibase 10.1086/497691} {\bibfield  {journal} {\bibinfo  {journal} {The
  Astrophysical Journal}\ }\textbf {\bibinfo {volume} {633}},\ \bibinfo {pages}
  {L65} (\bibinfo {year} {2005})}\BibitemShut {NoStop}%
\bibitem [{\citenamefont {Pace}\ and\ \citenamefont {Strigari}(2018)}]{jfact2}%
  \BibitemOpen
  \bibfield  {author} {\bibinfo {author} {\bibnamefont {Pace}, \bibfnamefont
  {A.~B.}}\ and\ \bibinfo {author} {\bibnamefont {Strigari}, \bibfnamefont
  {L.~E.}},\ }\href {\doibase 10.1093/mnras/sty2839} {\bibfield  {journal}
  {\bibinfo  {journal} {Mon. Not. R. Astr. Soc.}\ }\textbf {\bibinfo {volume}
  {482}},\ \bibinfo {pages} {3480} (\bibinfo {year} {2018})}\BibitemShut
  {NoStop}%
\bibitem [{\citenamefont {{Percival}}(2001)}]{wjp2001}%
  \BibitemOpen
  \bibfield  {author} {\bibinfo {author} {\bibnamefont {{Percival}},
  \bibfnamefont {W.~J.}},\ }\href {\doibase 10.1046/j.1365-8711.2001.04837.x}
  {\bibfield  {journal} {\bibinfo  {journal} {Mon. Not. R. Astr. Soc.}\
  }\textbf {\bibinfo {volume} {327}},\ \bibinfo {pages} {1313} (\bibinfo {year}
  {2001})},\ \Eprint {http://arxiv.org/abs/astro-ph/0107437}
  {arXiv:astro-ph/0107437 [astro-ph]} \BibitemShut {NoStop}%
\bibitem [{\citenamefont {Pieri}, \citenamefont {Bertone},\ and\ \citenamefont
  {Branchini}(2008)}]{Pieri_2008}%
  \BibitemOpen
  \bibfield  {author} {\bibinfo {author} {\bibnamefont {Pieri}, \bibfnamefont
  {L.}}, \bibinfo {author} {\bibnamefont {Bertone}, \bibfnamefont {G.}}, \ and\
  \bibinfo {author} {\bibnamefont {Branchini}, \bibfnamefont {E.}},\ }\href
  {\doibase 10.1111/j.1365-2966.2007.12828.x} {\bibfield  {journal} {\bibinfo
  {journal} {Mon. Not. R. Astr. Soc.}\ }\textbf {\bibinfo {volume} {384}},\
  \bibinfo {pages} {1627} (\bibinfo {year} {2008})},\ \Eprint
  {http://arxiv.org/abs/https://academic.oup.com/mnras/article-pdf/384/4/1627/2838096/mnras0384-1627.pdf}
  {https://academic.oup.com/mnras/article-pdf/384/4/1627/2838096/mnras0384-1627.pdf}
  \BibitemShut {NoStop}%
\bibitem [{\citenamefont {{Planck Collaboration}}, \citenamefont {{Aghanim}}\
  \emph {et~al.}(2018)\citenamefont {{Planck Collaboration}}, \citenamefont
  {{Aghanim}} \emph {et~al.}}]{planck}%
  \BibitemOpen
  \bibfield  {author} {\bibinfo {author} {\bibnamefont {{Planck
  Collaboration}},}, \bibinfo {author} {\bibnamefont {{Aghanim}}, \bibfnamefont
  {N.}},  \emph {et~al.},\ }\href@noop {} {\bibfield  {journal} {\bibinfo
  {journal} {arXiv e-prints}\ ,\ \bibinfo {eid} {arXiv:1807.06209}} (\bibinfo
  {year} {2018})},\ \Eprint {http://arxiv.org/abs/1807.06209} {arXiv:1807.06209
  [astro-ph.CO]} \BibitemShut {NoStop}%
\bibitem [{\citenamefont {{Press}}\ and\ \citenamefont
  {{Schechter}}(1974)}]{ex_dndm}%
  \BibitemOpen
  \bibfield  {author} {\bibinfo {author} {\bibnamefont {{Press}}, \bibfnamefont
  {W.~H.}}\ and\ \bibinfo {author} {\bibnamefont {{Schechter}}, \bibfnamefont
  {P.}},\ }\href {\doibase 10.1086/152650} {\bibfield  {journal} {\bibinfo
  {journal} {The Astrophysical Journal}\ }\textbf {\bibinfo {volume} {187}},\
  \bibinfo {pages} {425} (\bibinfo {year} {1974})}\BibitemShut {NoStop}%
\bibitem [{\citenamefont {Salucci}\ \emph {et~al.}(2012)\citenamefont {Salucci}
  \emph {et~al.}}]{burkert2}%
  \BibitemOpen
  \bibfield  {author} {\bibinfo {author} {\bibnamefont {Salucci}, \bibfnamefont
  {P.}} \emph {et~al.},\ }\href {\doibase 10.1111/j.1365-2966.2011.20144.x}
  {\bibfield  {journal} {\bibinfo  {journal} {Mon. Not. R. Astr. Soc.}\
  }\textbf {\bibinfo {volume} {420}},\ \bibinfo {pages} {2034} (\bibinfo {year}
  {2012})}\BibitemShut {NoStop}%
\bibitem [{\citenamefont {Sefusatti}\ \emph {et~al.}(2014)\citenamefont
  {Sefusatti}, \citenamefont {Zaharijas}, \citenamefont {Serpico},
  \citenamefont {Theurel},\ and\ \citenamefont
  {Gustafsson}}]{Sefusatti:2014vha}%
  \BibitemOpen
  \bibfield  {author} {\bibinfo {author} {\bibnamefont {Sefusatti},
  \bibfnamefont {E.}}, \bibinfo {author} {\bibnamefont {Zaharijas},
  \bibfnamefont {G.}}, \bibinfo {author} {\bibnamefont {Serpico}, \bibfnamefont
  {P.~D.}}, \bibinfo {author} {\bibnamefont {Theurel}, \bibfnamefont {D.}}, \
  and\ \bibinfo {author} {\bibnamefont {Gustafsson}, \bibfnamefont {M.}},\
  }\href {\doibase 10.1093/mnras/stu686} {\bibfield  {journal} {\bibinfo
  {journal} {Mon. Not. R. Astr. Soc.}\ }\textbf {\bibinfo {volume} {441}},\
  \bibinfo {pages} {1861} (\bibinfo {year} {2014})}\BibitemShut {NoStop}%
\bibitem [{\citenamefont {Servant}\ and\ \citenamefont
  {Tait}(2003)}]{servant_tait}%
  \BibitemOpen
  \bibfield  {author} {\bibinfo {author} {\bibnamefont {Servant}, \bibfnamefont
  {G.}}\ and\ \bibinfo {author} {\bibnamefont {Tait}, \bibfnamefont
  {T.~M.~P.}},\ }\href {\doibase 10.1016/S0550-3213(02)01012-X} {\bibfield
  {journal} {\bibinfo  {journal} {Nucl. Phys.}\ }\textbf {\bibinfo {volume}
  {B650}},\ \bibinfo {pages} {391} (\bibinfo {year} {2003})},\ \Eprint
  {http://arxiv.org/abs/hep-ph/0206071} {arXiv:hep-ph/0206071 [hep-ph]}
  \BibitemShut {NoStop}%
\bibitem [{\citenamefont {{Sheth}}, \citenamefont {{Mo}},\ and\ \citenamefont
  {{Tormen}}(2001)}]{Sheth_2001}%
  \BibitemOpen
  \bibfield  {author} {\bibinfo {author} {\bibnamefont {{Sheth}}, \bibfnamefont
  {R.~K.}}, \bibinfo {author} {\bibnamefont {{Mo}}, \bibfnamefont {H.~J.}}, \
  and\ \bibinfo {author} {\bibnamefont {{Tormen}}, \bibfnamefont {G.}},\ }\href
  {\doibase 10.1046/j.1365-8711.2001.04006.x} {\bibfield  {journal} {\bibinfo
  {journal} {Mon. Not. R. Astr. Soc.}\ }\textbf {\bibinfo {volume} {323}},\
  \bibinfo {pages} {1} (\bibinfo {year} {2001})},\ \Eprint
  {http://arxiv.org/abs/astro-ph/9907024} {arXiv:astro-ph/9907024 [astro-ph]}
  \BibitemShut {NoStop}%
\bibitem [{\citenamefont {{Sheth}}\ and\ \citenamefont
  {{Tormen}}(1999)}]{Sheth}%
  \BibitemOpen
  \bibfield  {author} {\bibinfo {author} {\bibnamefont {{Sheth}}, \bibfnamefont
  {R.~K.}}\ and\ \bibinfo {author} {\bibnamefont {{Tormen}}, \bibfnamefont
  {G.}},\ }\href {\doibase 10.1046/j.1365-8711.1999.02692.x} {\bibfield
  {journal} {\bibinfo  {journal} {Mon. Not. R. Astr. Soc.}\ }\textbf {\bibinfo
  {volume} {308}},\ \bibinfo {pages} {119} (\bibinfo {year} {1999})},\ \Eprint
  {http://arxiv.org/abs/astro-ph/9901122} {astro-ph/9901122} \BibitemShut
  {NoStop}%
\bibitem [{\citenamefont {{Stecker}}(1978)}]{Stecker:1978du}%
  \BibitemOpen
  \bibfield  {author} {\bibinfo {author} {\bibnamefont {{Stecker}},
  \bibfnamefont {F.~W.}},\ }\href {\doibase 10.1086/156336} {\bibfield
  {journal} {\bibinfo  {journal} {The Astrophysical Journal}\ }\textbf
  {\bibinfo {volume} {223}},\ \bibinfo {pages} {1032} (\bibinfo {year}
  {1978})}\BibitemShut {NoStop}%
\bibitem [{\citenamefont {Tavakoli}\ \emph {et~al.}(2014)\citenamefont
  {Tavakoli}, \citenamefont {Cholis}, \citenamefont {Evoli},\ and\
  \citenamefont {Ullio}}]{Tavakoli:2013zva}%
  \BibitemOpen
  \bibfield  {author} {\bibinfo {author} {\bibnamefont {Tavakoli},
  \bibfnamefont {M.}}, \bibinfo {author} {\bibnamefont {Cholis}, \bibfnamefont
  {I.}}, \bibinfo {author} {\bibnamefont {Evoli}, \bibfnamefont {C.}}, \ and\
  \bibinfo {author} {\bibnamefont {Ullio}, \bibfnamefont {P.}},\ }\href
  {\doibase 10.1088/1475-7516/2014/01/017} {\bibfield  {journal} {\bibinfo
  {journal} {JCAP}\ }\textbf {\bibinfo {volume} {1401}},\ \bibinfo {pages}
  {017} (\bibinfo {year} {2014})},\ \Eprint {http://arxiv.org/abs/1308.4135}
  {arXiv:1308.4135 [astro-ph.HE]} \BibitemShut {NoStop}%
\bibitem [{\citenamefont {Taylor}\ and\ \citenamefont
  {Silk}(2003)}]{Taylor:2002zd}%
  \BibitemOpen
  \bibfield  {author} {\bibinfo {author} {\bibnamefont {Taylor}, \bibfnamefont
  {J.~E.}}\ and\ \bibinfo {author} {\bibnamefont {Silk}, \bibfnamefont {J.}},\
  }\href {\doibase 10.1046/j.1365-8711.2003.06201.x} {\bibfield  {journal}
  {\bibinfo  {journal} {Mon. Not. R. Astr. Soc.}\ }\textbf {\bibinfo {volume}
  {339}},\ \bibinfo {pages} {505} (\bibinfo {year} {2003})}\BibitemShut
  {NoStop}%
\bibitem [{\citenamefont {Ullio}\ \emph {et~al.}(2002)\citenamefont {Ullio},
  \citenamefont {Bergstrom}, \citenamefont {Edsjo},\ and\ \citenamefont
  {Lacey}}]{Ullio:2002pj}%
  \BibitemOpen
  \bibfield  {author} {\bibinfo {author} {\bibnamefont {Ullio}, \bibfnamefont
  {P.}}, \bibinfo {author} {\bibnamefont {Bergstrom}, \bibfnamefont {L.}},
  \bibinfo {author} {\bibnamefont {Edsjo}, \bibfnamefont {J.}}, \ and\ \bibinfo
  {author} {\bibnamefont {Lacey}, \bibfnamefont {C.~G.}},\ }\href {\doibase
  10.1103/PhysRevD.66.123502} {\bibfield  {journal} {\bibinfo  {journal} {Phys.
  Rev.}\ }\textbf {\bibinfo {volume} {D66}},\ \bibinfo {pages} {123502}
  (\bibinfo {year} {2002})},\ \Eprint {http://arxiv.org/abs/astro-ph/0207125}
  {arXiv:astro-ph/0207125 [astro-ph]} \BibitemShut {NoStop}%
\bibitem [{\citenamefont {Wechsler}\ \emph {et~al.}(2002)\citenamefont
  {Wechsler}, \citenamefont {Bullock}, \citenamefont {Primack}, \citenamefont
  {Kravtsov},\ and\ \citenamefont {Dekel}}]{Wechsler_2002}%
  \BibitemOpen
  \bibfield  {author} {\bibinfo {author} {\bibnamefont {Wechsler},
  \bibfnamefont {R.~H.}}, \bibinfo {author} {\bibnamefont {Bullock},
  \bibfnamefont {J.~S.}}, \bibinfo {author} {\bibnamefont {Primack},
  \bibfnamefont {J.~R.}}, \bibinfo {author} {\bibnamefont {Kravtsov},
  \bibfnamefont {A.~V.}}, \ and\ \bibinfo {author} {\bibnamefont {Dekel},
  \bibfnamefont {A.}},\ }\href {\doibase 10.1086/338765} {\bibfield  {journal}
  {\bibinfo  {journal} {The Astrophysical Journal}\ }\textbf {\bibinfo {volume}
  {568}},\ \bibinfo {pages} {52–70} (\bibinfo {year} {2002})}\BibitemShut
  {NoStop}%
\end{thebibliography}%



\end{document}